\RequirePackage{ifpdf}
\documentclass[hyper,letterpaper]{JHEP3}
\usepackage{amsmath,amssymb,amsfonts,bm,amscd}
\usepackage{cite}
\usepackage{graphicx}
\usepackage{multirow}
\usepackage{verbatim}
\usepackage{appendix}
\usepackage{fancybox}
\usepackage{array}
\usepackage{url}
\usepackage{float}

\newcommand{\bea}{\begin{eqnarray}}
\newcommand{\eea}{\end{eqnarray}}
\newcommand{\be}{\begin{equation}}
\newcommand{\ee}{\end{equation}}

\newcommand{\R}{{\mathbb R}}

\def\Tr{{\rm Tr \,}}

\def\tilde{\widetilde}

\def\bar{\overline}

\def\CI{{\mathcal I}}

\def\CL{{\mathcal L}}

\def\CN{{\mathcal N}}

\newcommand{\cp}{{\mathbb{C}}{\mathbf{P}}}

\renewcommand{\bar}{\overline}

\newcommand{\circled}[1]{\text{\textcircled{\tiny $#1$}}}

\newcommand{\XX}{{\cal X}}
\newcommand{\YY}{{\cal Y}}
\newcommand{\ZZ}{{\cal Z}}

\title{$(0,2)$ Trialities}

\author{Abhijit Gadde$^{1}$, Sergei Gukov$^{1,2}$, and Pavel Putrov$^{1}$
\\
$^1$ California Institute of Technology, Pasadena, CA 91125, USA \\
$^2$ Max-Planck-Institut f\"ur Mathematik, Vivatsgasse 7, D-53111 Bonn, Germany}

\abstract{Motivated by the connection between 4-manifolds and 2d $\CN=(0,2)$ theories, we study the dynamics of a fairly large class of 2d $\CN=(0,2)$ gauge theories. We see that physics of such theories is very rich, much as the physics of 4d ${\cal N}=1$ theories. We discover a new type of duality that is very reminiscent of the 4d Seiberg duality. Surprisingly, the new 2d duality is an operation of order three: it is IR equivalence of three different theories and, as such, is actually a triality. We also consider quiver theories and study their triality webs. Given a quiver graph, we find that supersymmetry is dynamically broken unless the ranks of the gauge groups and flavor groups satisfy stringent inequalities. In fact, for most of the graphs these inequalities have no solutions. This supports the folklore theorem that generic 2d $\CN=(0,2)$ theories break supersymmetry dynamically.}

\preprint{CALT 68-2862}
\begin{document}
\cornersize{1}


\section{Introduction}

Recent years have seen the physics of gauge theories emerge from the M5 brane dynamics. When the M5 branes are compactified on a $d$-dimensional manifold $M_d$ with an appropriate partial topological twist, the physics in the remaining $6-d$ dimensions is expected to be described by a non-trivial superconformal field theory $T[M_d]$. Mapping $M_d$ to $T[M_d]$ becomes progressively harder as the dimension $d$ goes up. On the one hand, the world of $d$-manifolds becomes richer and wilder with larger values of $d=2,3,4, \ldots$ and, on the other hand, partial topological twist along $M_d$ leaves less and less supersymmetry in the remaining $6-d$ dimensions where $T[M_d]$ lives.

The program of analyzing $T[M_4]$ for general 4-manifolds was initiated in \cite{Gadde:2013sca}. The partial topological twist considered in \cite{Gadde:2013sca} leads to a 2d $\CN=(0,2)$ supersymmetric theory. In some simple cases, theories $T[M_4]$ labeled by 4-manifolds can be realized by a system of free (left-moving) fermions or their close cousins, such as $(0,2)$ coset models. However, in general, one needs to consider interacting gauge theories, such as variants of 2d $\CN=(0,2)$ SQED and SQCD. This makes the 4d-2d correspondence very interesting and challenging at the same time. We hope that pursing this program will benefit both fields and improve our understanding of 2d $\CN=(0,2)$ gauge theories as well as 4-manifolds. For example, it leads to a simple interpretation of Kirby moves as dualities in supersymmetric gauge theories and, in the opposite direction, predicts new dualities between 2d $\CN=(0,2)$ gauge theories that will be a starting point of our analysis here.

Even though 2d $\CN=(0,2)$ theories are of the utmost importance in constructing heterotic string models, surprisingly little is known about non-abelian gauge dynamics of 2d theories with $\CN=(0,2)$ supersymmetry. Ironically, there seem to be even more exact results about $\CN=0$ gauge theories with no supersymmetry that go back to the seminal work of 't Hooft \cite{'tHooft:1974hx}. Since in two dimensions the confinement is generic, even in abelian theories \cite{Schwinger:1962tp}, the effective physics is described by singlet states whose spectrum often can be determined exactly by large-$N$ techniques, bosonisation, or other methods. Also, a lot is known about models with larger $\CN=(2,2)$ supersymmetry, where additional constraints on dynamics allow to determine the IR fate of such theories. In contrast, very little is known about $(0,2)$ gauge dynamics, even with respect to the simplest abelian models like SQED.\footnote{It appears that non-abelian theories, such as 2d $\CN=(0,2)$ SQCD, have not been studied at all.} Part of the reason is that 2d $\CN=(0,2)$ theories often exhibit dynamical supersymmetry breaking and determining whether a given theory has SUSY vacua requires full-fledged analysis of quantum effects.

In this paper, we attempt to reduce this gap by studying non-abelian $\CN=(0,2)$ gauge theories in two dimensions. Such theories exhibit very rich dynamics and, as it turns out, enjoy interesting triality relations. This triality is similar in spirit to the Seiberg duality \cite{Seiberg:1994pq} of 4d ${\cal N}=1$ SQCD and, to the best of our knowledge, is the first example of a non-abelian gauge duality in 2d theories with $\CN=(0,2)$ supersymmetry.

The equivariant index ($a.k.a.$ the flavored elliptic genus) plays a key role in our analysis.
Although it has been extensively studied for $\CN = (2,2)$ NL$\sigma$/coset models, the tools
for computing it in gauge theories have been developed only recently \cite{Gadde:2013wq,Gadde:2013dda,Benini:2013nda,Benini:2013xpa}.
We use it to check the triality claim and also to learn about the low energy physics.
Most importantly, it serves as an excellent probe of dynamical supersymmetry breaking, which is essential in the study of 2d $\CN=(0,2)$ models.
As an aside, note that the $S^3\times S^1$ partition function (or, the ``Romelsberger index'')
can not be used to probe supersymmetry breaking in four dimensions.
It was pointed out in \cite{Festuccia:2011ws} that R-symmetry is needed in order to preserve supersymmetry on $S^3 \times S^1$.
However, unless the theory flows to a non-trivial fixed point, the R-symmetry is broken and the index simply doesn't make sense.
The 2d index is free of such demons because it is a partition function in flat space-time.

The outline of the rest of the paper is as follows. In section \ref{02theories}, we start by introducing the basics of $(0,2)$ gauge theories and analyze dynamical SUSY breaking in a prototype example of abelian model. Then, we gradually extend our analysis to more interesting gauge theories that were claimed to be dual to free fermions in \cite{Gadde:2013sca}. In section \ref{triality}, we consider the simplest but general non-abelian $(0,2)$ SQCD and formulate the triality proposal. The proposal is verified by matching the flavor symmetry anomalies, central charges, and the index. We study the low energy behavior as a function of ranks of flavor symmetry groups and give a general criterion for dynamical supersymmetry breaking. The fundamental SQCDs of section \ref{triality} are woven together to form complicated quivers in section \ref{trialitywebs}. We give general rules for triality transformations and study the triality webs in a few examples. We conclude the paper with an outlook in section \ref{outlook}.

\section{2d $\CN = (0,2)$ Gauge Theories}\label{02theories}

The $\CN = (0,2)$ supersymmetry in two dimensions admits three types of representations which are useful in constructing gauge theories. The first is the $(0,2)$ chiral multiplet (a.k.a. bosonic multiplet) $\Phi$. As the name suggests, it is annihilated by one of the superspace derivatives, $\bar {\cal D}_+ \Phi=0$, and has the expansion
\be
\Phi = \phi+\sqrt{2}\theta^{+}\psi_{+}-i\theta^{+}\bar{\theta}^{+}\partial_{+}\phi \,.
\ee
The chirality condition ensures that the component fermion $\psi_{+}$ is the right-moving one. The second multiplet is the $(0,2)$ Fermi multiplet $\Psi$. It obeys a similar condition, $\bar {\cal D}_+ \Psi=0$, that can be deformed to add an interaction with the chiral fields present in the theory, $\bar {\cal D}_+ \Psi_{a}=\sqrt{2} E_{a}(\Phi_{i})$. The components of the Fermi multiplet are
\be
\Psi=\psi_{-}-\sqrt{2}\theta^{+}G-i\theta^{+}\bar{\theta}^{+}\partial_{+}\psi_{-}-\sqrt{2}\bar{\theta}^{+}E \,.
\ee
The only on-shell degree of freedom is the left-moving fermion $\psi_{-}$.
In addition to the $E$-interaction, one can also add a superpotential term for the chiral and Fermi multiplets:
\be
\int d\theta^{+} \; \Psi_{a}J^{a}(\Phi_{i})|_{\bar{\theta}^{+}=0} \,.
\ee
Note that, unlike the superpotential in $\CN=(2,2)$ models, this is term is fermionic. The $E$-interaction can be exchanged for $J$-interaction at the expense of replacing the Fermi multiplet $\Psi$ with its conjugate multiplet $\bar{\Psi}$, which is also a Fermi multiplet. Supersymmetry requires the holomorphic $E_{a}$ and $J^{a}$ interactions to obey
\be
\sum_a E_{a}(\Phi_{i})J^{a}(\Phi_{i})=0 \,.
\ee
This condition is modified when 2d $\CN=(0,2)$ theory is realized on the boundary of 3d $\CN=2$ theory with a non-trivial superpotential \cite{Gadde:2013sca}.

The last and the most important ingredient of the gauge theory is the $(0,2)$ vector multipet. It is a real superfield with the expansion
\be
V=v-2i\theta^{+}\lambda_{-}-2i\bar{\theta}^{+}\bar{\lambda}_{-}+2\theta^{+}\bar{\theta}^{+}D \,.
\ee
The gauge invariant field strength belongs to a Fermi multiplet $\Lambda$.
The Fayet-Illiopoulos term is added to the gauge theory as $\frac{t}{4}\int d\theta^{+}\Lambda|_{\bar{\theta}^{+}=0}$,
where $t\equiv ir+\frac{\theta}{2\pi}$ combines the FI parameter and the $\theta$-angle.

As an example, we write down the Lagrangian of an abelian $(0,2)$ gauge theory
with chiral multiplets $\Phi_{i}$ of charge $q_i$ and Fermi multiplets $\Psi_{a}$ of charge $q_a$:
\be
{\cal L} = {\cal L}_{\text{gauge}}+{\cal L}_{\Phi}+{\cal L}_{\Psi}+{\cal L}_{FI}+{\cal L}_{J}
\label{tree-level}
\ee
where
\begin{eqnarray}
& {\cal L}_{\text{gauge}} & = \frac{1}{8e^{2}}\int d^{2}\theta \; \bar{\Lambda}\Lambda
\; = \; \frac{1}{e^{2}}\Big(\frac{1}{2}F_{01}^{2}+i\bar{\lambda}_{-}\partial_{+}\lambda_{-}+\frac{1}{2}D^{2}\Big) \nonumber\\
& {\cal L}_{\Phi} & = -\frac{i}{2}\int d^{2}\theta \; \bar{\Phi}_{i} \nabla_{-} \Phi^{i} \nonumber\\
& & = -|D_{\mu}\phi_{i}|^{2} + i\bar{\psi}_{+i} D_{-} \psi_{+}^{i}
- \sqrt{2}iq_i \bar{\phi}_{i} \lambda_{-} \psi_{+}^{i}
+ \sqrt{2}iq_i \phi^{i} \bar{\psi}_{+i} \bar{\lambda}_{-} + q_i |\phi_{i}|^{2}D \nonumber\\
& {\cal L}_{\Psi} & = -\frac{1}{2}\int d^{2}\theta \; \bar{\Psi}_{a} \Psi^{a} \nonumber\\
& & = i\bar{\psi}_{-a} D_+ \psi_{-}^{a} + |G_{a}|^{2} - |E_{a}(\phi)|^{2}
- \bar{\psi}_{-a} \frac{\partial E^{a}}{\partial\phi_{i}}\psi_{+i}
-\frac{\partial\bar{E}_{a}}{\partial\bar{\phi}_{i}}\bar{\psi}_{+i}\psi_{-}^{a} \nonumber\\
& {\cal L}_{FI} & = \frac{t}{4} \int d\theta^{+} \; \Lambda|_{\bar{\theta}_{+}=0}+c.c. \; = \; -rD+\frac{\theta}{2\pi}F_{01} \nonumber\\
& {\cal L}_{J} & = \int d\theta^{+} \; \Psi_{a}J^{a}(\Phi)|_{\bar{\theta}_{+}=0}+c.c.
\; = \; \sqrt{2}G_{a}J^{a}(\phi)+\psi_{-a}\psi_{+i}\frac{\partial J^{a}}{\partial\phi^{i}}+c.c.\nonumber
\end{eqnarray}
After eliminating the auxiliary fields the potential for the scalars $\phi_{i}$ is
\be
V \; = \; \frac{e^{2}}{2}\Big(\sum_{i}q_{i}|\phi_{i}|^{2}-r\Big)^{2} + \sum_a |E_a (\phi)|^{2} + \sum_a |J_a (\phi)|^{2} \,.
\ee
In order for the gauge theory to make sense at the quantum level, we should make sure that the gauge anomaly is zero. It is given by
\be
\Tr \gamma^3 G G \; = \; \sum_{i:~\text{chiral}} q_i^2 - \sum_{a:~\text{Fermi}} q_a^2
\ee
where ``$G$'' stands for ``Gauge'' here and in what follows .

\subsection{Warm-up: a $(0,2)$ deformation of $\cp^{N-1}$ model}

Let us analyze quantum aspects of a concrete example in more detail:  a $(0,2)$ deformation of the $\cp^{N-1}$ sigma-model realized as a gauged linear sigma-model (GLSM). After studying dynamical supersymmetry breaking we then add various bells and whistles to this model, eventually constructing a large class of new 2d superconformal theories with $\CN=(0,2)$ supersymmetry as well as new dual pairs.

Specifically, our starting point is a 2d $\CN=(0,2)$ gauged linear sigma model with $U(1)$ gauge group and the following matter fields:
\be
\begin{array}{l@{\;}|@{\;}ccc}
& ~\Sigma & ~~\Phi_{i=1, \ldots, N} & ~~\Psi_{i=1, \ldots, N} \\\hline
U(1)_{\text{gauge}}~ & ~0 & ~~+1 & ~~+1
\end{array}
\ee
where $\Sigma = \sigma + \sqrt{2}\theta^{+} \lambda_{+} - \ldots$ and $\Phi_i$ are $(0,2)$ chiral multiplets, while $\Psi_i$ are Fermi multiplets. Note, this theory has no gauge anomaly since it contains equal number of $(0,2)$ chiral and Fermi multiplets of charge $+1$. We also include in this $(0,2)$ model a holomorphic $E$-interaction
\be
E_j = i \epsilon \sqrt{2} \Sigma \Phi_j.
\label{EtermCPNmodel}
\ee
that modifies the chirality constraint $\bar {\cal D}_+ \Psi_j = \sqrt{2} E_j$ for each Fermi multiplet and will play a crucial role in what follows. In particular, we wish to analyze the role of this interaction, as a function of the parameter $\epsilon$, on the dynamical supersymmetry breaking. Note, this theory interpolates between $\CN=(2,2)$ gauged linear sigma-model (when $\epsilon = 1$) and a $\CN=(0,2)$ model with free chiral multiplet $\Sigma$ (when $\epsilon = 0$).

The Lagrangian \eqref{tree-level} also includes a Fayet-Iliopoulos (FI) term with complex coefficient $t = ir + \tfrac{\theta}{2\pi}$:
\be
\CL_{FI} = \frac{t}{4} \int \theta^+ \Lambda \vert_{\bar \theta^+ = 0} + c.c. = - rD + \frac{\theta}{2\pi} F_{01}
\ee
{}From the experience with the $(2,2)$ locus, we know that the dependence of the bare Fayet-Iliopoulos parameter on the UV cut-off $\Lambda_{UV}$ is
\be
r_0 = N \log \left( \frac{\Lambda_{UV}}{\mu} \right)
\ee

Our next goal is to analyze the dynamics of this theory. Following \cite{Witten:1978bc} (see also \cite{Tong:2007qj,Shifman:2008kj,Bolokhov:2010hv}), we consider the large-$N$ approximation which amounts to evaluating one-loop determinants of charged matter fields. Integrating out $\Phi_i$ and $\Psi_i$ can be done in superspace, keeping $\CN=(0,2)$ supersymmetry manifest \cite{McOrist:2007kp}. The result is the effective Lagrangian for the superfields $\Lambda$ and $\Sigma$ that, besides the terms already present in \eqref{tree-level}, also contains a 1-loop contribution:
\be
\CL_{\tilde J} = \int d \theta^+ \; \Lambda \tilde J (\Sigma) \vert_{\bar \theta^+ = 0} + c.c.
\label{L1loop}
\ee
which has the form of a field-dependent FI term and plays the role of a ``twisted superpotential'' in a 2d theory with $\CN=(0,2)$ supersymmetry \cite{Gadde:2013sca}. In the $(0,2)$ deformation of the $\cp^{N-1}$ linear sigma-model considered here the Coulomb branch is parametrized by the vev of $\Sigma$ that makes $\Phi_i$ and $\Psi_i$ massive. Specifically, from \eqref{EtermCPNmodel} we see that the mass matrix is a $N \times N$ matrix with all eigenvalues equal to $\epsilon \Sigma$. Therefore, evaluating the determinant of this matrix we find
\be
\tilde J = \frac{i}{8 \pi} \log \frac{(\epsilon \sigma)^N}{q \mu^N}
\ee
where $q = e^{2 \pi i t (\mu)}$.
Hence, we conclude that for generic values of $\epsilon \ne 0$ the theory has $N$ massive supersymmetric vacua at
\be
\sigma^N = \frac{q \mu^N}{\epsilon^N}
\ee
which are deformations of the $N$ vacua in the familiar $\cp^{N-1}$ sigma-model with $\CN=(2,2)$ supersymmetry. In the limit $\epsilon \to 0$ these vacua run off to infinity indicating dynamical SUSY breaking of the minimal $(0,2)$ model.

It is instructive to write the interaction \eqref{L1loop} in components:
\be
\CL_{\tilde J} = - 4 \text{Im} (\tilde J) D + 4 \text{Re} (\tilde J) F_{01}
- 8 i \frac{\partial \tilde J}{\partial \sigma} \lambda_- \lambda_+
+ 8 i \frac{\partial \overline{ \tilde J}}{\partial \bar \sigma} \bar \lambda_- \bar \lambda_+
\ee
If we also knew the 1-loop correction to the kinetic terms in the Lagrangian \eqref{tree-level},
we could consistently compute the effective scalar potential for the fields $\sigma$ and $D$.
Unfortunately, such a 1-loop computation does not seem to be available in the literature.
However, one might hope to reproduce qualitative features of the effective scalar potential
by using the tree-level kinetic terms, which yield
\be
V_{\text{eff}} (\sigma, D) = \frac{1}{2 e^2} D^2 - rD
- \frac{N}{2 \pi} D \log \vert \frac{\epsilon \sigma}{\mu} \vert
\label{Veffa}
\ee
Indeed, this scalar potential leads to the same conclusion --- namely, that our theory has massive SUSY vacua for non-zero values of $\epsilon$ and dynamical SUSY breaking for $\epsilon = 0$ --- but now we can see a little more directly how and why this happens.
It would be interesting to study loop corrections to the kinetic terms.
Relegating this problem to future work, we can compare the structure of \eqref{Veffa}
with the effective scalar potential computed in the large-$N$ approximation, as in \cite{Shifman:2008kj,Bolokhov:2010hv}.
In this approach, the analogue of the last term in \eqref{Veffa}
comes from evaluating one-loop determinants of charged matter fields\footnote{From here on, all dimensionful quantities are written in units of the coupling constant $e$.}
\be
\prod_{i=1}^{N} \det ((\partial_{\mu} + i A_{\mu})^2 + |\epsilon \sigma|^2)
\ee
in the case of $N$ Dirac fermions and, similarly,
\be
\prod_{i=1}^{N} \frac{1}{\det ((\partial_{\mu} + i A_{\mu})^2 - D + |\epsilon \sigma|^2)}
\ee
in the case of $N$ charged scalars.
Note, this ratio of one-loop determinants exhibits the standard boson-fermion cancelation in the supersymmetric vacuum with $D=0$.

Another important feature of these one-loop determinants is that the auxiliary field $D$ appears only in the denominator ({\it i.e.} only in the scalar field contribution). The reason for this is that in the tree-level Lagrangian \eqref{tree-level} the field $D$ only affects the mass matrix of scalar fields, but not the fermions. Moreover, the contribution of $D$ to the mass of a given scalar field is proportional to its charge. This is a general fact that holds even in models without $\Sigma$ field (that we are going to consider shortly).

Therefore, we learn that one simple way to ensure that SUSY is {\it not} dymanically broken in a general $(0,2)$ model with charged chiral and Fermi multiplets is to consider equal number of {\it chiral} multiplets with positive and negative charge. Even in models without $\Sigma$-field(s) and the corresponding $E$-terms, this will guarantee that $V_{\text{eff}} (D)$ is an even function of $D$, {\it i.e.} has a critical point at $D=0$. (In fact, it is easy to check that, in such cases, $D=0$ is a minimum with $V_{\text{eff}}=0$.)

Before we proceed to more general theories, let us point out that in the limit $\epsilon = 0$ the effective potential $V_{\text{eff}} (D)$ only depends on $D$ and not $\sigma$ (since $\Sigma$ is free in this limit). In particular, evaluating the above determinants it is easy to see that  $V_{\text{eff}} (D)$ has the critical point at
\be
ir + N \int \frac{d^2 k}{(2\pi)^2} \frac{1}{k^2 - D} = 0
\ee
leading to the SUSY breaking expectation value
\be
D^N = 4^N \Lambda^{2N} \equiv 4^N \mu^{2N} e^{4\pi i t}
\ee
On the other hand, modifying the mass matrix by the $E$-terms $E_i = M_{ij} \Phi_j$ changes the critical point of the effective potential to
\be
\det \left( M^{\dagger} M + D \cdot {\bf 1}_{N \times N} \right) = 4^N \Lambda^{2N}
\ee
which does restore supersymmetry at the appropriately tuned value of $M$.
The general conclusion of this analysis is that incorporating superpotential terms often helps to avoid dynamical supersymmetry breaking
in this class of 2d $\CN=(0,2)$ models.
This conclusion is certainly consistent with the earlier study of $(0,2)$ models \cite{DistlerKachru,Silverstein:1994ih}
and will be a useful guide to us in what follows.

Although we have given semiclassical arguments for the supersymmetry breaking in the limit $\epsilon \to 0$, perhaps the strongest support for these claims comes from the computation of the elliptic genus. The elliptic genus is a (refined) Witten index of the theory quantized on a circle. Therefore a non-zero elliptic genus indicates that the supersymmetry is unbroken dynamically. We will see that the elliptic genus of the theory with $\epsilon \neq 0$ is non-zero while it vanishes for $\epsilon =0$. This holds even for the case of finite $N$. Before getting into this analysis let us take a slight detour and review the machinery necessary to compute the elliptic genus.

\subsection*{The elliptic genus}

Recently there has been some progress in computing the elliptic genus of the 2d gauge theory. In \cite{Gadde:2013wq}, the authors discussed elliptic genus of $\CN=(0,2)$ gauge theory, while a prescription for computing $\CN=(2,2)$ elliptic genus was given in \cite{Gadde:2013dda} motivated by the Gauss law. In \cite{Benini:2013nda,Benini:2013xpa}, the $\CN=(0,2)$ as well as  $\CN=(2,2)$ elliptic genus was derived from rigorous path integral localization. We will summarize the prescription for a general $\CN=(0,2)$ gauge theories below. A reader interested in the derivation is encouraged to look at the references cited above.

The elliptic genus is simplest to define in radial quantization:
\be
{\cal I}(a_i;q) \; = \; \Tr (-1)^F q^{L_0} \prod_i a_i^{f_i} \,.
\ee
For convenience we take the Hilbert space to be in the NS-NS sector. Only the states satisfying the NS shortening condition ${\bar L}_0 =\frac{1}{2} {\bar J}_0$ contribute to the index \footnote{We will use the terms `elliptic genus' and `index' interchangeably.}. We have refined the usual definition of the elliptic genus by adding the fugacities $a_i$ that keep track of all flavor symmetries. A chiral multiplet and a Fermi multiplet whose primary has $J_0=R$ contribute, respectively,
\be\label{multipletindex}
{\cal I}_{\Phi} = \theta(q^{\frac{R}{2}} a;q)^{-1}
\qquad\text{and}\qquad
{\cal I}_{\Psi} = \theta(q^{\frac{R+1}{2}} a;q) \,.
\ee
Here $a$ is the fugacity that is associated to a $U(1)$ symmetry that acts on these multiplets. Here, we introduced $\theta (a;q)=(a;q)(q/a;q)$ and $(a;q)=\prod_{i=0}^{\infty}(1-aq^i)$. Only the gauge invariant degrees of freedom of the vector multiplet, {\it i.e.} its field strength multiplet $\Lambda$, contributes to the index. For the $U(1)$ case, ${\cal I}_{\Lambda}^{U(1)}=(q;q)^2$, and for $G=U(N)$:
\be
{\cal I}_{\Lambda}^{U(N)} \; = \; (q;q)^{2N}\prod_{i\neq j}\theta (a_i/a_j;q) \,.
\ee
Here $a_i, i=1,\ldots,N$ are the fugacities associated to the Cartan generators of the $U(N)$ gauge group.
Then, the index of a general 2d $\CN=(0,2)$ theory is computed by the following prescription:
\begin{enumerate}
\item Multiply the contribution of all the multiplets while keeping track of the flavor symmetries. Thanks to the gauge anomaly cancellation this is an elliptic function of the gauge fugacities.
\item Evaluate the residues at the poles in the fundamental domain  coming from  positively (or negatively) charged chiral multiplets.
\end{enumerate}

We are now ready to compute the elliptic genus of the $\cp^{N-1}$ model and its $(0,2)$ deformation. The index of the $\cp^{N-1}$ model is given by
\be
{\cal I}\; = \; (q;q)^2 \oint \frac{dz}{2\pi i z} \frac{1}{\theta(x;q)} \prod_{i=1}^{N}\frac {\theta (q x^{-1} a_i/z;q)}{\theta (z/a_i;q)} \,.
\ee
In addition to the gauge fugacity $z$ and $SU(N)$ flavor fugacities $a_i$ (s.t. $\prod a_i=1$), we have also introduced the fugacity $x$ for the $U(1)$ symmetry acting on the neutral chiral field $\Sigma$ and the Fermi fields $\Psi_i$. When we shift $z\to qz$, the integrand gets multiplied by $x^N$. It is an elliptic function of $z$ only when $x^N=1$. This indicates that quantum mechanically the $U(1)_x$ symmetry is broken to ${\mathbb Z}_N$. Evaluating the residues at $z=a_j$, we get
\be
{\cal I} \; = \; \sum_j \prod_{i\neq j}\frac{\theta(x a_j/a_i)}{\theta(a_j/a_i)} \,.
\ee
When we set $x=1$, we see that the index is $N$. This allows us to conclude that the supersymmetry is unbroken for the $\cp^{N-1}$ model and that it in fact has $N$ vacua. When we get rid of the $\Sigma$ field and the superpotential, the first term in the integrand disappears. Also the non-abelian flavor symmetry enhances to $SU(N)\times SU(N)$ with each $SU(N)$ acting on $N$ Fermi and $N$ chiral multiplets separately. We introduce new $SU(N)$ fugacities $b_i$. Evaluating the residues, we get
\be
{\cal I} \; = \; (q;q)^2 \oint \frac{dz}{2\pi i z} \prod_{i=1}^{N}\frac {\theta (q x^{-1} b_i/z;q)}{\theta (z/a_i;q)}
\; = \; \sum_j \theta (x a_j/b_j)\prod_{i\neq j}\frac{\theta(x a_j/b_i)}{\theta(a_j/a_i)} \,.
\ee
It is quite non-trivial, but this expression does vanish for $x^N=1$. We have checked this analytically for $N=2$ and in $q$-expansion for higher $N$.

\subsection{Superconformal theories from 4-manifolds}

In \cite[sec. 3.5]{Gadde:2013sca}, the authors found new 2d $\CN=(0,2)$ superconformal field theories that are expected to be dual to theories of free fermions. These dualities were motivated by gluing operations on 4-manifolds. In this section we will revisit these theories and analyze them in detail. Later we will see that these theories can be generalized to a much larger class which have nontrivial fixed points and exhibit even more interesting dualities.

\subsection*{Abelian}

The simplest example of the 2d $\CN=(0,2)$ theory encountered in \cite{Gadde:2013sca} that is dual to free fermions is the abelian gauge gauge theory with one chiral multiplet $\Phi$ of charge $1$ and $N_f$ Fermi multiplets $\Psi_i$ of charge $-1$.
This theory, as it stands, has gauge anomaly that can be canceled by integrating in $N_f-1$ pairs of chiral and Fermi multiplets $(P_a,\Gamma_a)$ where $P_a$ has gauge charge $-1$ and $\Gamma_a$ is neutral:
\be
\begin{array}{l@{\;}|@{\;}cccc}
& ~~\Phi & ~~~~\Psi_{i=1, \ldots, N_f} & ~~~~P_{a=1, \ldots, N_f-1} & ~~~~\Gamma_{a=1, \ldots, N_f-1} \\\hline
U(1)_{\text{gauge}}~ & ~~+1 & ~~~~-1 & ~~~~-1 & ~~~~0
\end{array}
\ee
These fields are coupled via a $J$-term superpotential
\be
{\cal L}_J \; = \; \int d\theta^+ \; \Phi P_a \Gamma_a |_{{\bar \theta}^+=0} \,.
\ee
Classically, the $D$-term equation in this model has the form
\be
r - |\phi^2| + \sum_{a=1}^{N_f-1} |p_a|^2 \; = \; 0
\ee
and quantum mechanically (if we are in the regime $N_f \ge 2$) the value of $r$
is renormalized to the ``large volume region,'' thus forcing $\phi$ to get a vev.
When $\phi$ gets a vev, the anomaly-canceling pairs $(P_a, \Gamma_a)$ all become massive
and can be integrated out in a manifestly $\CN=(0,2)$ supersymmetric way,
leading to the ``twisted superpotential'' \eqref{L1loop} with
\be
\tilde J \; = \; - \frac{i}{8 \pi} (N_f - 1) \log (\Phi)
\label{tJPGout}
\ee
This is precisely the ``charged log interaction'' of \cite{Melnikov:2012nm}, which, in fact, was introduced precisely as a result of integrating out massive pairs $(P_a, \Gamma_a)$ with unbalanced charge. The resulting low-energy theory now contains one $(0,2)$ chiral superfield $\Phi$ of charge $+1$ and $N_f$ Fermi multiplets $\Psi_i$ of charge $-1$ coupled to the gauge multiplet $\Lambda$:
\be
\begin{array}{l@{\;}|@{\;}cc}
& ~\Phi & ~~~\Psi_{i=1, \ldots, N_f}  \\\hline
U(1)_{\text{gauge}}~ & ~+1 & ~~~-1
\end{array}
\ee
The chiral anomaly in this model is canceled against the ``classical anomaly'' ({\it i.e.} gauge non-invariance) of the term \eqref{tJPGout}. Including the contribution of this term, the effective scalar potential for the fields $D$ and $\phi$ then takes the form:
\be
V_{\text{eff}} (D,\phi) \; = \; \frac{1}{2} D^2 + D \left( |\phi|^2 - r + \frac{N_f - 1}{2\pi} \log |\phi| \right)
\ee
This potential has a supersymmetric minimum (with $D=0$) at
\be
|\phi|^2 + \frac{N_f - 1}{2\pi} \log |\phi| \; = \; r
\ee
for all values of $r$, including $r=0$. Again, $N_f \ge 2$ turns out to be a crucial condition for this, and $\phi$ getting a vev justifies integrating out the pairs $(P_a, \Gamma_a)$.

The R-symmetry at low energies is typically different from the canonical R-symmetry. When possible, it can be determined by imposing the cancellation of the mixed anomaly with the gauge symmetry. In two-dimensional $\CN=(0,2)$ theories with $E$-term and $J$-term interactions, the low-energy R-symmetry was studied in \cite{Silverstein:1994ih}.
Without sufficiently many superpotential couplings, however, the R-charge may not be pinned down uniquely. This phenomenon is similar to the one in four dimensions, where the R-symmetry is determined by the principle of ``a-maximization'' \cite{Intriligator:2003jj}. In two dimensions, the corresponding quantity is the central charge that, according to the Zamolodchikov's $c$-theorem \cite{Zamolodchikov:1986gt}, wants to decrease (and, in fact, was part of the motivation for the ``a-maximization'' \cite{Intriligator:2003jj}). Its extremization in 2d $(0,2)$ theories was implemented in \cite{Benini:2013cda}, where it was shown that in a model with normalizable vacuum state the low-energy R-symmetry extremizes $c_R$. This condition is equivalent to the condition of vanishing mixed anomaly with \emph{all} abelian symmetries. In particular, this means that the mixed anomaly with the gauge symmetry automatically vanishes.

The superconformal symmetry relates the right-moving central charge $c_R$ to the anomaly in R-symmetry. The left-moving central charge $c_L$ can then  be computed using $c_R$ and the gravitational anomaly:
\be\label{central-charges}
c_R = 3\Tr \gamma^3 RR \,,\qquad c_R-c_L=\Tr \gamma^3 \,.
\ee
In the model of interest,
\be
\frac{c_R}{3} \; = \; (R_\Phi -1)^2 - R_{\Psi}^2 N_f + (R_P-1 )^2 (N_f-1)-R_\Gamma^2 (N_f-1)-1 \,.
\ee
The last term is the contribution from the vector multiplet. The trial central charge needs to be extremized subject to the superpotential constraint $R_\Phi + R_P+R_\Gamma=1$. We get
\be
R_\Phi=R_\Psi=0 \,,\qquad R_P=\frac{N_f-2}{N_f-1} \,,\qquad R_\Gamma=\frac{1}{N_f-1} \,.
\ee
The central charges for these values of R-charge are $(c_L , c_R) = (N_f , 0)$. They are consistent with our proposal that this theory is dual to a theory of $N_f$ free Fermi multiplets $\Gamma'$. Note, each Fermi multiplet contributes $\Delta c_L = 1$ to the left-moving central charge and does not contribute to the right-moving central charge $c_R$. The duality proposal is summarized in the table below.
\begin{center}
~~~~~~~~~~~~~~~~~~2d~$\CN=(0,2)$~SQED~~~~~~~~~~~~~~~~~~~~~~~~~~~~free~fermions \\[.2cm]
\begin{tabular}{c|cccccccccc|c}
 & $\Phi$ & $\Psi$ & $P$ & $\Gamma$ &  &  &  &  &  &  & $\Gamma'$\tabularnewline
\cline{1-5} \cline{12-12}
$U(1)_{\mbox{gauge}}$ & $+1$ & $-1$ & $-1$ & $0$ &  &  &  &  &  &  & \tabularnewline
$SU(N_{f})$ & $ $${\bf 1}$ & $\square$ & $ $${\bf 1}$ & $ $${\bf 1}$ &  &  & $\quad\simeq$ &  &  &  & $\square$\tabularnewline
$SU(N_{f}-1)$ & ${\bf 1}$ & ${\bf 1}$ & $\square$ & $\overline{\square}$ &  &  &  &  &  &  & ${\bf 1}$\tabularnewline
\end{tabular}
\par\end{center}
One can easily calculate the anomalies of the non-abelian symmetry. On the gauge theory side,
\begin{eqnarray}
\Tr \, \gamma^3 J_{SU(N)} J_{SU(N)} & = & -T_{\Psi}(\square) \; = \; -\frac{1}{2} \,, \\
\Tr \, \gamma^3 J_{SU(N_f-1)} J_{SU(N_f-1)} & = & T_{P}(\square)-T_{\Gamma}({\overline \square}) \; = \; 0 \,. \nonumber
\end{eqnarray}
The non-abelian anomalies of Free fermions are precisely the same.
Physically, the dual Fermi multiplets are the gauge invariant mesonic operators
\be
\Gamma' \; = \; \Phi \Psi \,.
\ee
Our analysis of R-symmetries shows that both $\Phi$ and $\Psi$ have canonical R-charges in the infra-red and do not develop any anomalous dimensions. This is the reason why the mesonic operators $\Phi \Psi$ also have the canonical R-charge and can be described by free Fermi multiplets.

We can present a strong evidence for this duality by computing the elliptic genus (where, on the gauge theory side, we use the superconformal R-charges determined above). Using the basic ingredients \eqref{multipletindex} we get
\begin{eqnarray}
{\cal I} & = & (q;q)^{2}\oint\frac{dz}{2\pi iz}\frac{1}{\theta(z)}\prod_{i=1}^{N_{f}}\theta(q^{\frac{1}{2}}x_{i}/z)\left(\prod_{a=1}^{N_{f-1}}\frac{\theta(q^{\frac{1}{2}(1+\frac{1}{N_{f}-1})}s_{a}^{-1})}{\theta(q^{\frac{1}{2}(1-\frac{1}{N_{f}-1})}s_{a}/z)}\right)\nonumber\\
 & = & \prod_{i=1}^{N_{f}}\theta(q^{\frac{1}{2}}x_{i}) \,.
\end{eqnarray}
The contribution of the $(P,\Gamma)$ pair is shown in the brackets in the first line. They neatly cancel when we evaluate the residue, giving us the index of $N_{f}$ free fermions.

\subsection*{Non-abelian}

In \cite[sec. 3.5]{Gadde:2013sca}, the authors also found a non-abelian version of the duality. It involves a $U(N_{c})$ gauge theory with $N_{c}$ chiral multiplets $\Phi^{\alpha}_s$ in the fundamental representation and $N_{f}$ Fermi multiplets $\Psi_{\alpha}^i$ in the anti-fundamental representation. Here $\alpha=1,\ldots ,N_c$ is the color label, $s=1,\ldots, N_c$ is the $SU(N_c)'$ flavor label\footnote{The prime on $SU(N_c)'$ just serves to distinguish the flavor symmetry from the $SU(N_c)$ part of the gauge symmetry.} and $i=1,\ldots, N_f$ is the $SU(N_f)$ flavor label.

The chiral and Fermi multiplets contribute $\frac{1}{2}N_{c}$ and $-\frac{1}{2}N_{f}$ to the $SU(N_c)$ gauge anomaly, respectively. The non-abelian vector multiplet itself contributes $-N_{c}$ to the gauge anomaly, resulting in the net anomaly of $-\frac{1}{2}(N_{f}+N_{c})$. As before, this anomaly can be canceled by introducing $N_{f}+N_{c}$ chiral-Fermi pairs $(P^{a}_\alpha,\Gamma_{a}^s)$, where only $P^{a}$ transforms as the anti-fundamental while $\Gamma_{a}$ is neutral under gauge symmetry. The label $a=1,\ldots, N_c+N_f$ is the $SU(N_f+N_c)$ flavor symmetry label. In addition to the $SU(N_c)$ part of the gauge symmetry, we also need to cancel the anomaly for the $U(1)$ part. To that effect we introduce two extra Fermi multiplets $\Omega_{1,2}$ in the determinant representation.\footnote{If one chooses to work with the $SU(N_c)$ gauge group, then there is no need to add the extra $\Omega$ multiplets. The $U(1)$ symmetry would then be a baryonic flavor symmetry of the theory. In the rest of the paper, we will consider only the $U(N_c)$ gauge theory.}

The theory has a $J$-term interaction
\be\label{J-term}
{\cal L}_J \; = \; \int d\theta^+ \; \Phi^{\alpha}_s P^{a}_{\alpha} \Gamma_{a}^s|_{{\bar \theta}^+=0}
\ee
as in the abelian case. This theory is claimed to be dual to the theory of $N_{c}N_{f}+2$ free fermions ${\Gamma'}^{i}_s$ and $\Omega'_{1,2}$. The gauge and flavor charges of all the fields are summarized in the table below.
\begin{center}
~~~~~~~~~~~~~~~~~~2d~$\CN=(0,2)$~SQCD~~~~~~~~~~~~~~~~~~~~~~free~fermions \\[.2cm]
\begin{tabular}{c|cccccc|cc}
 & $\Phi$ & $\Psi$ & $P$ & $\Gamma$ & $\Omega$ &  & $\Gamma'$ & $\Omega'$\tabularnewline
\cline{1-6} \cline{8-9}
$U(N_{c})$ & $\square$ & $\overline{\square}$ & $\overline{\square}$ & $ $${\bf 1}$ & $\mbox{det}$ &  & $ $$ $ & $ $\tabularnewline
$SU(N_{f})$ & $ $${\bf 1}$ & $\square$ & $ $${\bf 1}$ & $ $${\bf 1}$ & ${\bf 1}$ & $\qquad\simeq\qquad$ & $\square$ & ${\bf 1}$\tabularnewline
$SU(N_{c})'$ & $\overline{\square}$ & ${\bf 1}$ & ${\bf 1}$ & ${\square}$ & ${\bf 1}$ &  & $\overline{\square}$ & ${\bf 1}$\tabularnewline
$SU(N_{f}+N_{c})$ & ${\bf 1}$ & ${\bf 1}$ & $\square$ & $\overline{\square}$ & ${\bf 1}$ &  & ${\bf 1}$ & ${\bf 1}$\tabularnewline
$SU(2)$ & ${\bf 1}$ & ${\bf 1}$ & ${\bf 1}$ & ${\bf 1}$ & $\square$ &  & ${\bf 1}$ & $\square$\tabularnewline
\end{tabular}
\par\end{center}

The trial central charge in this case is,
\be
\frac{c_R}{3}= N_c^2(R_\Phi-1)^2-N_cN_f R_\Psi^2+N_c(N_c+N_f)(R_P-1)^2-N_c(N_c+N_f)R_\Gamma^2-2R_\Omega^2+\frac{c_R(G)}{3}\nonumber
\ee
Here $c_R(G)=-3N_c^2$ is a fixed contribution  from the vector multiplet. It doesn't play any role in determining the superconformal R-charges.  Extremizing subject to the superpotential relation $R_\Phi+R_P+R_\Gamma=1$, we get
\be
R_\Phi=R_\Psi=R_\Omega=0,\qquad R_P=\frac{N_f}{N_f+N_c},\qquad R_\Gamma=\frac{N_c}{N_f+N_c}.
\ee
At these values of the R-charge we find $c_R=0$. This matches the central charge of the dual theory because free Fermi multiplets do not contribute to the right-moving central charge. We also get $c_L=N_f N_c+2$ which matches with the total number of Fermi multiplets on the dual side. We can also match the flavor anomalies as we did in the abelian case. On the gauge theory side,
\begin{eqnarray}
\mbox{Tr}[\gamma^{3}J_{SU(N_{f})}J_{SU(N_{f})}] & = & T_{\Psi}(\square)N_{c}=-\frac{N_{c}}{2}\\
\mbox{Tr}[\gamma^{3}J_{SU(N_{f}+N_{c})}J_{SU(N_{f}+N_{c})}] & = & [T_{P}(\square)N_{c}-T_{\Gamma}(\overline{\square})N_{c}]=0\\
\mbox{Tr}[\gamma^{3}J_{SU(N_{c})'}J_{SU(N_{c})'}] & = & [T_{P}(\square)N_{c}-T_{\Gamma}(\overline{\square})(N_{f}+N_{c})]=-\frac{N_{f}}{2} \,.
\end{eqnarray}
It is very easy to see that the anomaly contribution of the system of $N_{f} N_{c}$ fermions transforming as $(\square,\overline{\square})$ under $SU(N_{f})\times SU(N_{c})'$ is exactly same as above. The $\Omega'$ fermions do not contribute to these anomalies. Finally, we support our claim by showing the equality of the index on both sides of the proposed duality:
\begin{eqnarray}
{\cal I} & = & (q;q)^{2}\oint\prod_{\alpha=1}^{N_{c}}\frac{d\xi_{\alpha}}{2\pi i\xi_{\alpha}}\prod_{\alpha\neq\beta}\theta(\frac{\xi_{\alpha}}{\xi_{\beta}})
\frac{\prod_{\alpha,i}\theta(q^{\frac{1}{2}}z_{i}\xi_{\alpha}^{-1})}{\prod_{\alpha,s}\theta(\xi_{\alpha}d_{s})}
\theta(q^{\frac{1}{2}}w \prod_\alpha \xi_\alpha)\theta(q^{\frac{1}{2}}w^{-1} \prod_\alpha \xi_\alpha)
\nonumber\\
&\times &
\left(\frac{\prod_{s,a}\theta(q^{1-\frac{1}{2}\frac{N_{f}}{N_{c}+N_{f}}}d_{s}^{-1}c_{a}^{-1})}{\prod_{\alpha,a}\theta(q^{\frac{1}{2}\frac{N_{f}}{N_{c}+N_{f}}}c_{a}\xi_{\alpha}^{-1})}\right) \nonumber \\
 & = & \theta(q^{\frac{1}{2}}w)\theta(q^{\frac{1}{2}}w^{-1})\prod_{i,s}\theta(q^{\frac{1}{2}}z_{i}d_{s}) \,.
\end{eqnarray}
We see that the integral is precisely the index of $N_fN_c+2$ free Fermi multiplets. Just as before, the dual fermions $\Gamma'$ can be also thought of as the mesonic operators $\Phi \Psi$ of the electric theory. Again, because $\Phi$ and $\Psi$ have canonical R-charges in the infra-red, the meson corresponds to a free field.

The gauge theory considered here is dual to the theory of only free mesons. This is strongly reminiscent of the 4d ${\cal N}=1$ SQCD with $N_f=N_c$ or $N_f=N_c+1$. It is then natural to look for the analogue of the Seiberg duality in SQCD with general values of $N_f$. In the next section we will consider such a generalization and will be pleasantly surprised by the result.

\section{The Fundamental Triality}\label{triality}

\subsection{Proposal}

\begin{figure}[ht]
\centering
\includegraphics[scale=0.8]{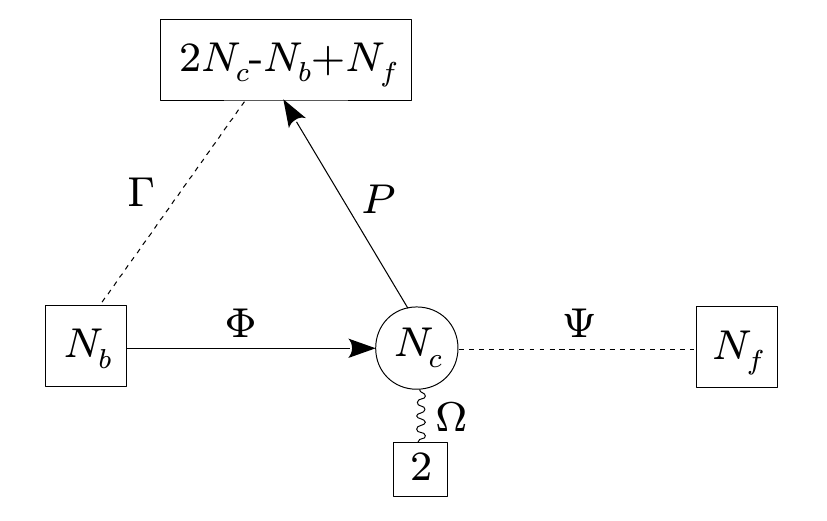}
\caption{The $(0,2)$ SQCD. We use oriented solid arrows to label chiral fields with their representations, while unoriented dotted lines represent Fermi multiplets. The $\Omega$ multiplet in the $\mbox{det}$ representation is shown with a wavy line.}
\label{fig:qcd}
\end{figure}

Consider a $U(N_c)$ gauge theory but now with $N_b$ fundamental chiral multiplets and $N_f$ anti-fundamental Fermi multiplets. The $SU(N_c)$ anomaly cancellation condition requires that we add $2N_c+N_f-N_b$ chiral multiplets $P$ in the anti-fundamental representation. We also add the same number of Fermi fields $\Gamma$ that transform in the fundamental of $SU(N_b)$ flavor symmetry. All in all, the field content is the same as before except that $SU(N_c)'$ is generalized to  $SU(N_b)$ and $SU(N_f+N_c)$ is generalized to $SU(2N_c+N_f-N_b)$.  For convenience, it is summarized below:
\begin{center}
\begin{tabular}{c|ccccc}
\multicolumn{6}{c}{~~2d~$\CN=(0,2)$~SQCD}  \tabularnewline[.2cm]
 & $\Phi$ & $\Psi$ & $P$ & $\Gamma$ & $\Omega$\tabularnewline
\hline
$U(N_{c})$ & $\square$ & $\overline{\square}$ & $\overline{\square}$ & $ $${\bf 1}$ & $\mbox{det}$\tabularnewline
$SU(N_{f})$ & $ $${\bf 1}$ & $\square$ & $ $${\bf 1}$ & $ $${\bf 1}$ & ${\bf 1}$\tabularnewline
$SU(N_{b})$ & $\overline{\square}$ & ${\bf 1}$ & ${\bf 1}$ & ${\square}$ & ${\bf 1}$\tabularnewline
$SU(2N_{c}+N_{f}-N_{b})$ & ${\bf 1}$ & ${\bf 1}$ & $\square$ & $\overline{\square}$ & ${\bf 1}$\tabularnewline
$SU(2)$ & ${\bf 1}$ & ${\bf 1}$ & ${\bf 1}$ & ${\bf 1}$ & $\square$\tabularnewline
\end{tabular}
\par\end{center}
We listed here all flavor symmetries of the theory except two $U(1)$ symmetries; they will be discussed in section \ref{checks}. The field content allows us to write the superpotential $J=\Phi P \Gamma$. The gauge theory can be neatly represented in terms of a quiver diagram in figure \ref{fig:qcd}. The superpotential term is associated to the closed triangular loop in the quiver diagram.

Motivated by 4d Seiberg duality, we expect to find a dual theory which is a $U(N_b-N_c)$  gauge theory. The bilinear fields $\Phi\Psi$ are expected to be the mesonic fields in the dual theory. They should transform in the bi-fundamental of the $SU(N_b)\times SU(N_f)$ flavor symmetry. Moreover, they should couple to the ``magnetic" matter multiplets $\Phi'$ and $\Psi'$ through the cubic superpotential. But such superpotential is impossible to write down as it is not fermionic. Also, if we require only $\Phi'$ and $\Psi'$ to be charged under the dual gauge group, the gauge anomaly is not cancelled unless they are equal in number. This clearly  presents a problem in matching the flavor symmetries on dual side. As we will see momentarily, these problems neatly cancel each other and we get an elegant and symmetric proposal for the duality if we introduce the chiral fields $P'$:
\begin{center}
Proposed dual $(0,2)$ SQCD:$\hspace{1em}\qquad$%
\begin{tabular}{cc|ccccc}
 &  & $\Phi'$ & $\Psi'$ & $P'$ & $\Gamma'$ & $\Omega'$\tabularnewline
\cline{2-7}
 & $U(N_{b}-N_{c})$ & $\square$ & $\overline{\square}$ & $\overline{\square}$ & $ $${\bf 1}$ & $\mbox{det}$\tabularnewline
 & $SU(2N_c+N_f-N_b)$ & $ $${\bf 1}$ & $\square$ & $ $${\bf 1}$ & $ $${\bf 1}$ & ${\bf 1}$\tabularnewline
 & $SU(N_{f})$ & $\overline{\square}$ & ${\bf 1}$ & ${\bf 1}$ & ${\square}$ & ${\bf 1}$\tabularnewline
 & $SU(N_b)$ & ${\bf 1}$ & ${\bf 1}$ & $\square$ & $\overline{\square}$ & ${\bf 1}$\tabularnewline
 & $SU(2)$ & ${\bf 1}$ & ${\bf 1}$ & ${\bf 1}$ & ${\bf 1}$ & $\square$\tabularnewline
\end{tabular}
\par\end{center}
Examining the representations of matter fields it is easy to see that this duality not only changes the rank of the gauge group as in Seiberg duality of 4d $\CN=1$ theories but also permutes the three flavor symmetries:
\bea
N_f & \mapsto & 2N_c + N_f - N_b  \nonumber\\
N_b & \mapsto &  N_f\label{duality-map} \\
2N_c+N_f-N_b & \mapsto & N_b \nonumber
\eea
Let us define this transformation as $D$. It is consistent with the change in the rank of the gauge group $N_c \mapsto N_b-N_c$.

Moreover, in the original theory, the roles of $\Phi$ and $P$ are exchanged under charge conjugation. Of course the charge conjugation is not a symmetry of the theory but one can conjugate, dualize and conjugate back to get a yet another dual description of the original theory. The rank of the gauge group in this description is going to be $(2N_c+N_f-N_b)-N_c=N_c+N_f-N_b$. A more algebraic way to obtain this new description is to observe that the transformation \eqref{duality-map}, unlike most of the ``dualities",  has order $3$. Hence we call it a triality. Application of $D$ and $D^2$ to the $U(N_c)$ gauge theory leads to $U(N_b-N_c)$ and $U(N_c+N_f-N_b)$ gauge theories, respectively.

\begin{figure}[ht]
\centering
\includegraphics[scale=0.8]{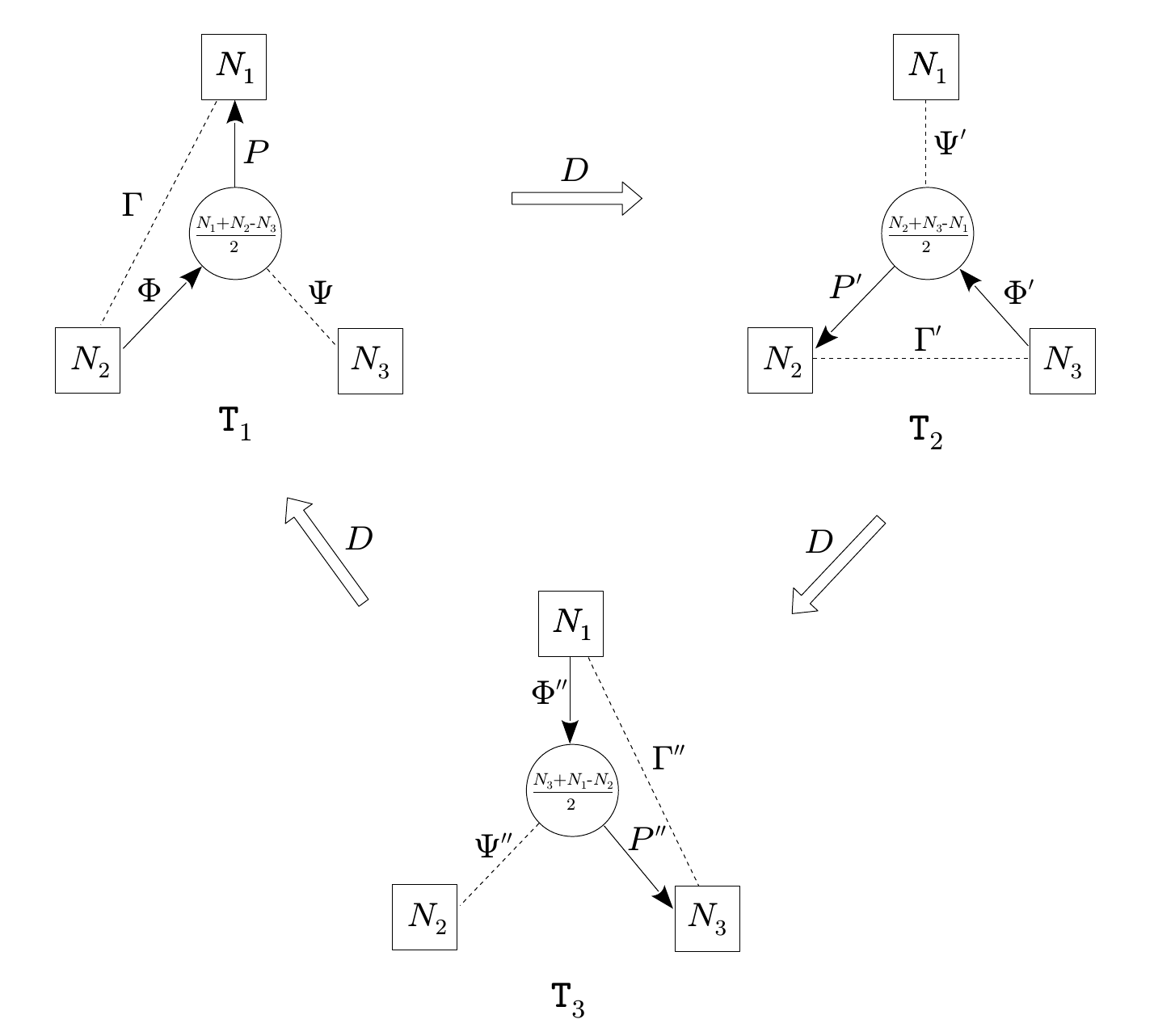}
\caption{The 2d $\CN=(0,2)$ triality. In this and the following figures the $\Omega$ multiplets are  suppressed.}
\label{fig:triality}
\end{figure}

In order to make the triality manifest, it is best to take the flavor symmetry groups to be $SU(N_1),SU(N_2)$ and $SU(N_3)$. The 2d $\CN=(0,2)$ triality is summarized in figure \ref{fig:triality}.

\subsection{Checks}\label{checks}
We now  support our proposal by matching the anomalies, central charges, and elliptic genera of dual theories. The flavored (a.k.a. equivariant) elliptic genus is a powerful quantity. As we show towards the end of  appendix \ref{appendix-theta}, it can used to read off all the anomalies of the theory including central charges. Nevertheless, we will compute the anomalies explicitly and show that they are the same in all duality frames. Note that it suffices to compute these quantities in one duality frame, say ${\tt T}_1$, and check that they are symmetric under the cyclic permutations of $N_1,N_2$ and $N_3$.

\subsection*{Non-abelian flavor anomalies}

Let us start with the simplest check, {\it i.e.} matching of non-abelian flavor anomalies. The $SU(N)$ anomalies of the ${\tt T}_1$ theory are
\bea
\Tr \gamma^3 J_{SU(N_1)} J_{SU(N_1)} &=& \frac{N_1+N_2-N_3}{2} T_{P} ({\square})-N_2 T_\Gamma(\overline\square)=-\frac{1}{4}(-N_1+N_2+N_3)\\
\Tr \gamma^3 J_{SU(N_2)} J_{SU(N_2)} &=& \frac{N_1+N_2-N_3}{2} T_{\Phi} (\overline {\square})-N_1 T_\Gamma({\square})=-\frac{1}{4}(+N_1-N_2+N_3)\\
\Tr \gamma^3 J_{SU(N_3)} J_{SU(N_3)} &=& -\frac{N_1+N_2-N_3}{2} T_{\Psi}({\square})=-\frac{1}{4}(+N_1+N_2-N_3).
\eea
Indeed, these expressions are invariant under cyclic permutations of $N_1,N_2$ and $N_3$. In addition to these, we have a $SU(2)$ symmetry acting on the $\Omega$ Fermi multiplets. It is clear that its anomaly is the same in all duality frames.

\subsection*{Central charges}
Next, we determine the R-charge using c-extremization and compute the central charges $c_R$ and $c_L$. The trial central charge is
\be\nonumber
\frac{c_R}{3}=\frac{N_1+N_2-N_3}{2}((R_P-1)^2N_1-R_\Psi^2 N_3+(R_\Phi-1)^2N_2)-R_\Gamma^2N_1 N_2-2R_\Omega^2+\frac{c_R(G)}{3}
\ee
The term $c_R(G)=-3N_c^2=-\frac{3}{4}(N_1+N_2-N_3)^2$ is a fixed contribution from the $U(N_c)$ vector multiplet.  This is because FI term is linear in the field strength multiplet and has a fixed R-charge equal to $1$. Extremization of the trial $c_R$ gives us
\bea
R_\Phi & = & \frac{N_2+N_3-N_1}{N_1+N_2+N_3},\qquad R_\Psi=R_\Omega=0 \\
R_P & = & \frac{N_1-N_2+N_3}{N_1+N_2+N_3},\qquad R_\Gamma=\frac{N_1+N_2-N_3}{N_1+N_2+N_3}.
\eea
With these R-charges, using \eqref{central-charges}, we get
\bea
c_R & = & \frac{3}{4}\frac{(-N_1+N_2+N_3) (N_1-N_2+N_3) (N_1+N_2-N_3)}{N_1+N_2+N_3}\\
c_L & = & c_R- \frac{1}{4}(N_1^2+N_2^2+N_3^2-2N_1N_2-2N_2 N_3-2N_3N_1)+2.
\eea
Remarkably, both $c_R$ and $c_L$ are invariant under the permutations of $(N_1,N_2,N_3)$. This serves as a strong check of the proposed triality.

\subsection*{Abelian symmetry}
The gauge theory we are interested in has two abelian flavor symmetries that we call $F$ and $\tilde F$. We propose the following action on the matter fields:

\begin{center}
\begin{tabular}{c|c|c|c|c|c}
 & $\Phi$ & $\Psi$ & $P$ & $\Gamma$ & $\Omega$\tabularnewline
\hline
$F$ & $N_3-N_2$ & $0$ & $N_2-N_1$ & $N_1-N_3$ & $\frac{1}{2}(N_1 N_3-N_2^2)$\tabularnewline
$\tilde F$ & $N_2-N_1$ & $0$ & $N_1-N_3$ & $N_3-N_2$ & $\frac{1}{2}(N_2 N_3-N_1^2)$\tabularnewline
\end{tabular}
\par\end{center}
Their anomalies can be computed in a straightforward way. We get
\bea
\Tr \gamma^3 F^2 & = &\frac{1}{2}N_1 N_2 N_3 (N_1+N_2+N_3-\frac{N_1^2}{N_3}-\frac{N_2^2}{N_1}-\frac{N_3^2}{N_2})\\
\Tr \gamma^3 {\tilde F}^2 & = &\frac{1}{2}N_1 N_2 N_3 (N_1+N_2+N_3-\frac{N_1^2}{N_2}-\frac{N_2^2}{N_3}-\frac{N_3^2}{N_1})\\
\Tr \gamma^3 F\tilde F & = &-\frac{1}{2}N_1 N_2 N_3 (N_1+N_2+N_3-\frac{N_1 N_2}{N_3}-\frac{N_2 N_3}{N_1}-\frac{N_3 N_1}{N_2})
\eea
As we can see, the anomaly matrix of the two $U(1)$ symmetries is invariant under the cyclic permutations of $(N_1,N_2,N_3)$.

\subsection*{Index}
In this section we will compute the equivariant index of the theory in description ${\tt T}_1$. We use the fugacities $\{y_a: a=1,\ldots N_1\}$, $\{x_s: s=1,\ldots N_2\}$, and $\{z_i: i=1,\ldots N_3\}$ for the flavor symmetry groups $SU(N_1)$, $SU(N_2)$, and $SU(N_3)$, respectively. They satisfy $\prod x_s=\prod y_i=\prod z_a=1$. For the $U(N_c)$ gauge symmetry, we use the fugacity $\{\zeta_\alpha:\alpha=1,\ldots, N_c\}$. The fugacity $w$ is used for the $SU(2)$ that acts on the $\Omega$ multiplets. To avoid clutter, we will not introduce any fugacities for the $U(1)$ symmetries $F$ and $\tilde F$. Then, the index of the SQCD is
\be\label{SQCDindex}
{\cal I}  =  (q;q)^{2N_{c}}\oint\prod_{\alpha=1}^{N_{c}}\frac{d\zeta_{\alpha}}{2\pi i\zeta_{\alpha}}\prod_{\alpha\neq\beta}\theta(\zeta_{\alpha}/\zeta_{\beta})\frac{\prod_{a,s}\theta(q^{\frac{1+R_{\Gamma}}{2}}x_{s}/y_{a})\prod_{\alpha,i}\theta(q^{\frac{1}{2}}z_{i}/\zeta_{\alpha})\prod_{\pm}\theta(q^{\frac{1}{2}}w^{\pm}\prod_{\alpha}\zeta_{\alpha})}{\prod_{\alpha,s}\theta(q^{\frac{R_{\Phi}}{2}}\zeta_{\alpha}/x_{s})\prod_{\alpha,a}\theta(q^{\frac{R_{P}}{2}}y_{a}/\zeta_{\alpha})}
\ee
where the contour integral should be understood as sum over the residues at leading poles, either coming from the contribution of $\Phi$ or from the contribution of $P$. Let us pick the former set of poles. The simultaneous poles in all $N_c$ variables $\zeta_\alpha$ are classified by injective map $\sigma:\{\zeta_\alpha\}\to \{x_s\}$. Letting $\{\tilde x_\alpha\}$ to be the image of this map, the poles are at $\zeta_\alpha=q^
{-\frac{R_\Phi}{2}} \tilde x_\alpha$. Evaluating the residue,
\be
{\cal I}=\sum_{\{\tilde{x}_{\alpha}\}\subset\{x_{s}\}}\prod_{\alpha\neq\beta}\theta(\tilde{x}_{\alpha}/\tilde{x}_{\beta})\frac{\prod_{a,s}\theta(q^{\frac{1+R_{\Gamma}}{2}}x_{s}/y_{a})\prod_{\alpha,i}\theta(q^{\frac{1+R_{\Phi}}{2}}z_{i}/\tilde{x}_{\alpha})\prod_{\pm}\theta(q^{\frac{1-N_{c}R_{\Phi}}{2}}w^{\pm}\prod_{\alpha}\tilde{x}_{\alpha})}{\prod_{\alpha,a}\theta(q^{\frac{R_P-R_{\Phi}}{2}}\tilde{x}_{\alpha}/y_{a})\prod_{x_{s}\neq\tilde{x}_{\alpha}}\theta({\tilde x}_{\alpha}/x_{s})}.
\ee
This expression can be rewritten in terms of the variables $\{{\bar x}_{\bar \alpha}\}\equiv \{x_s\} \setminus \{ {\tilde x}_\alpha\}$. After some manipulations, we get
\be\label{new-residue}
{\cal I}=\sum_{\{\bar{x}_{\bar{\alpha}}\}\subset\{x_{s}\}}
\prod_{\bar{\alpha}\neq\bar{\beta}}\theta(\bar{x}_{\bar{\alpha}}/{\bar x}_{\bar{\beta}})\frac{\prod_{s,i}\theta(q^{\frac{1-R_{\Phi}}{2}}x_{s}/z_{i})
\prod_{\bar{\alpha},a}\theta(q^{\frac{1+R_{\Gamma}}{2}}{{\bar x}_{\bar{\alpha}}}/y_{a})\prod_{\pm}\theta(q^{\frac{1+N_{c}R_{\Phi}}{2}}w^{\pm}\prod_{\bar{\alpha}}\bar{x}_{\bar{\alpha}})}{\prod_{\bar{\alpha},i}\theta(q^{\frac{R_{P}+R_{\Gamma}}{2}}\bar{x}_{\bar{\alpha}}/z_i)\prod_{x_{s}\neq\bar{x}_{\bar{\alpha}}}\theta(x_s/\bar{x}_{\bar{\alpha}})}.
\ee
In writing this expression we used the theta function identity  $\theta(a)=\theta (q/a)$ and the superpotential constraint  $R_\Phi+R_P+R_\Gamma=1$. The expression \eqref{new-residue} is precisely the residue of
\be
{\cal I}=(q;q)^{2{\tilde N}_c}\oint\prod_{\bar{\alpha}=1}^{{\tilde N}_c}\frac{d\xi_{\bar{\alpha}}}{2\pi i\xi_{\bar{\alpha}}}\prod_{\bar{\alpha}\neq\bar{\beta}}\theta(\xi_{\bar{\alpha}}/\xi_{\bar{\beta}})
\frac{
\prod_{s,i}\theta(q^{\frac{1+R_{\Phi}}{2}}z_i/x_{s})
\prod_{\bar{\alpha},a}\theta(q^{\frac{1}{2}}\xi_{\bar{\alpha}}/y_{a})
\prod_{\pm}\theta(q^{\frac{1}{2}}w^{\pm}\prod_{\bar{\alpha}}\xi_{\bar{\alpha}})}
{
\prod_{\bar{\alpha},i}\theta(q^{\frac{R_{P}}{2}}\xi_{\bar{\alpha}}/z_i)
\prod_{\alpha,s}\theta(q^{\frac{R_{\Gamma}}{2}}x_s/\xi_{\bar{a}})}
\ee
where ${\tilde N}_c=N_2-N_c=(N_2+N_3-N_1)/2$. This is exactly the index of the dual theory  ${\tt T}_2$ where $\xi_{\bar \alpha}$ plays the role of the gauge fugacity. This is because $R_{P'}=R_\Gamma$, $R_{\Gamma'}=R_{\Phi}$, $R_{\Phi'}=R_{P}$, and also $N_c R_\Phi= {\tilde N}_c R_\Gamma$.

\subsection{Phase diagram}

\begin{figure}[ht]
\centering
\includegraphics[scale=0.3]{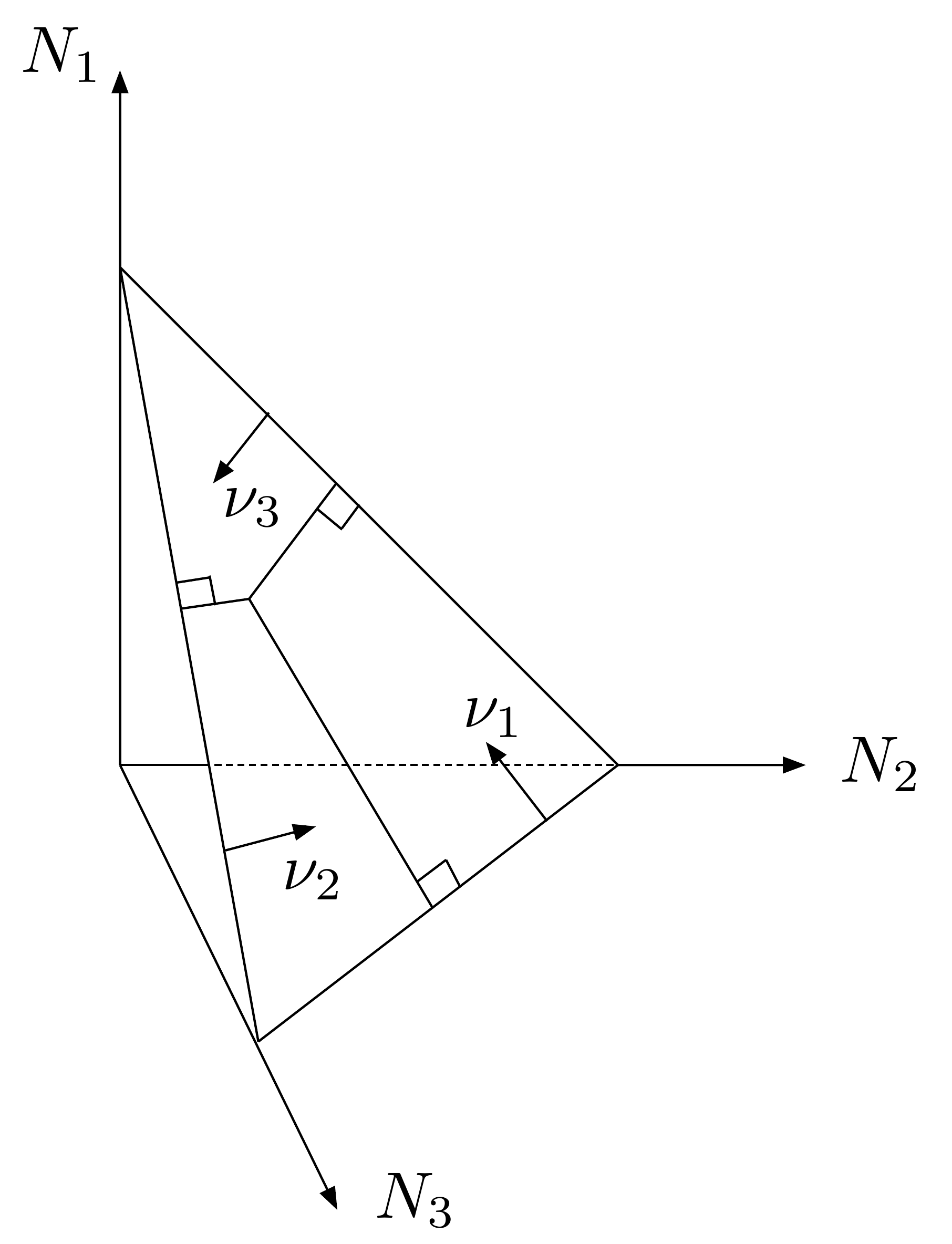}
\caption{The space of UV SQCDs. The triangular slice is the projective space that labels the theories up to a simultaneous rescaling of all $N_i$. Each edge of this slice has size $\tfrac{2}{\sqrt 3}$ and the center of mass coordinates $\nu_i$ are the distances of a given interior point from the three edges.}
\label{fig:parameterspace}
\end{figure}

In this section we  analyze the low energy physics of the 2d $\CN=(0,2)$ SQCD as a function of $N_i$ up to an overall rescaling $N_i \to \alpha N_i$. This parameter space  is best described in terms of the ``center of mass'' coordinates  $\nu_i \equiv \frac{N_i}{\sum_j N_j}$, which have the property $\nu_i\geq 0$ and $\sum  \nu_i=1$. They parametrize a solid equilateral triangle with sides $\tfrac{2}{\sqrt 3}$ shown in figure \ref{fig:parameterspace}, which is the space of all UV SQCDs upto an overall rescaling of $N_i$.

\begin{figure}[ht]
\centering
\includegraphics[scale=0.3]{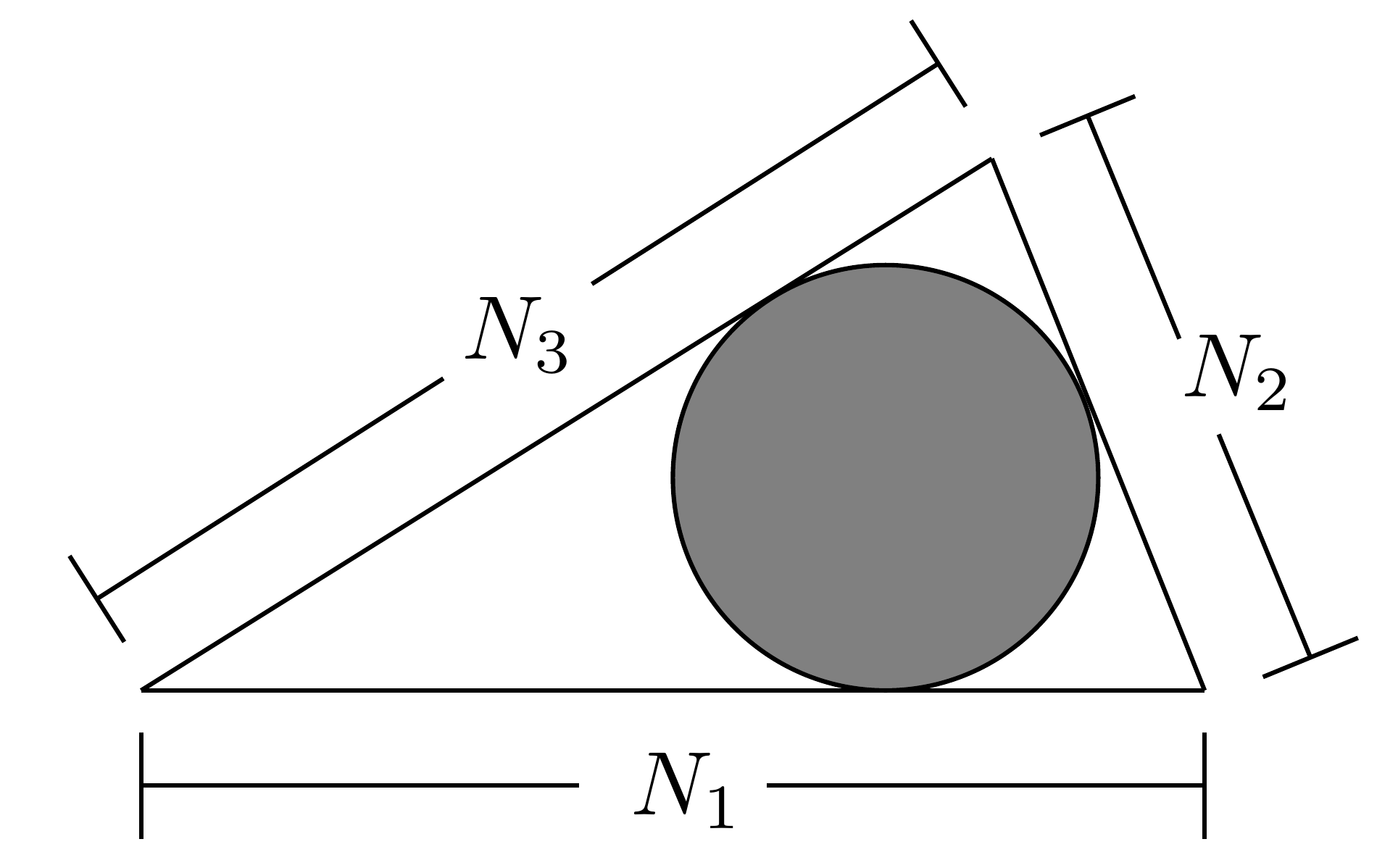}
\caption{The triangle labeling the SQCD. The area of the inscribed circle is equal to $\frac{\pi}{3} c_R$.}
\label{Ntriangle}
\end{figure}

In the last section the equivariant index provided us with a powerful check of the triality; in this section we will see that it is very useful in understanding the infra-red physics as well. First thing to notice is that if $N_2< N_c$, {\it i.e.} $N_2+N_3<N_1$, the integral \eqref{SQCDindex} does not admit any poles. The index is simply zero. This strongly suggests that supersymmetry is dynamically broken when  $N_2+N_3<N_1$. Applying the same argument in all duality frames, we come up with two more inequalities that signal the dynamical supersymmetry breaking: $N_1+N_2<N_3$ and $N_1+N_3<N_2$. Indeed, it is precisely when one of these these inequalities is  satisfied, there exists a duality frame in which the rank of the gauge group is negative.
This leads us to conclude that the supersymmetry is dynamically broken unless the $N_i$'s satisfy the triangle inequality. Figure \ref{Ntriangle} represents a typical SQCD. Curiously, the area of the inscribed circle is equal to $\frac{\pi}{3}c_R$.
The triangle inequality carves out a smaller equilateral triangle in the projective space parametrized by $\nu$'s. This smaller triangle has sides of size $\tfrac{1}{\sqrt 3}$ and represents the space of all  SQCDs that preserve supersymmetry in the IR. The triality acts on this space by a $\frac{2\pi}{3}$ rotation.

The triangle of $\nu_i$'s degenerates on the boundary of the supersymmetric parameter space. For example, when $\nu_1=\nu_2+\nu_3$ the rank of the gauge group is zero in the duality frame ${\tt T}_2$, {\it i.e.} the SQCD is actually dual to a theory of free fermions. This is also the case for all the theories corresponding to boundary points. At the corners of the parameter space, things degenerate even further. As an  example, consider the vertex with $\nu_3=0$ and $\nu_1=\nu_2$. In descriptions ${\tt T}_2$ and ${\tt T}_3$ this actually corresponds to the theory consisting of only two Fermi multiplets $\Omega$.  One can explicitly verify it by showing that the index of the gauge theory in description ${\tt T}_1$ is product of two $\theta$ functions. Even though the  ${\tt T}_1$ description consists of a non-trivial gauge theory, the index tells us that the low energy theory consists of only two left-moving fermionic degrees of freedom.

\begin{figure}[t]
\centering
\includegraphics[scale=0.3]{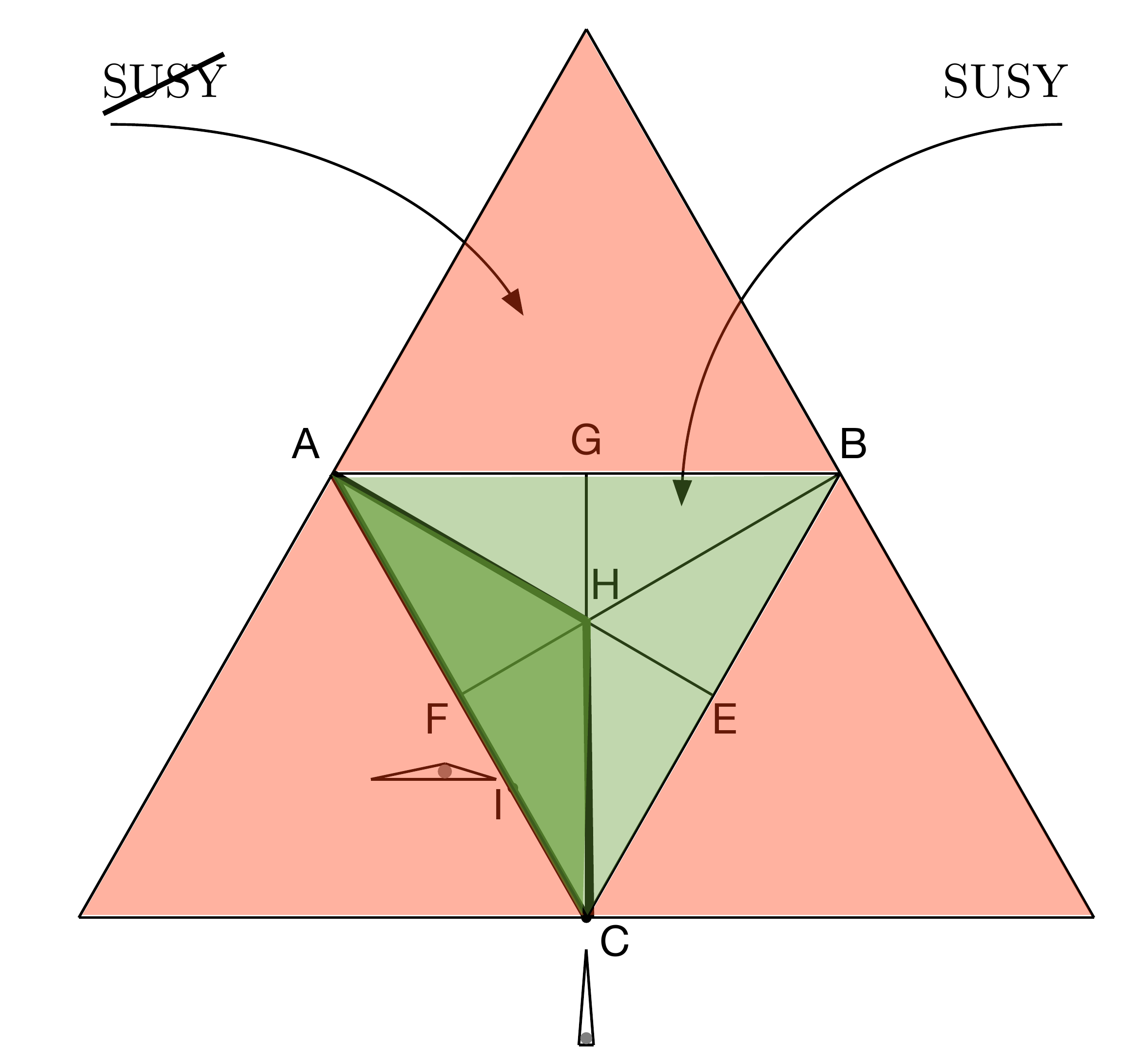}
\caption{The green triangle $ABC$ is the space of theories preserving supersymmetry, while the points in red correspond to SUSY breaking theories. The triality acts as a $\frac{2\pi}{3}$ rotation. The triangle $AHC$ shaded in dark green is the fundamental domain under the action of the triality. The points on the edges correspond to degeneration of the $N_i$ triangle in figure \protect\ref{Ntriangle}. The small triangle at $F$ denotes a typical degeneration. The corresponding theories are dual to free fermion theories. The $N_i$ triangle degenerates even further at the vertices of $ABC$. See the small triangle at $C$ for an example. The corresponding theories are empty in the infra-red except for two Fermi multiplets. The theories on the segments $AE$, $BF$ and $CG$ are expected to have exactly marginal deformations. The point $H$ is invariant under the triality. It correspond to the theory with $N_1=N_2=N_3$.
}
\label{fig:phasediagram}
\end{figure}

Another special locus is when the $\nu_i$ triangle becomes isosceles. Let us take $\nu_1=\nu_2$ as an example. In description ${\tt T}_3$, this theory has equal number of $\Phi$s and $P$s. These fields are charged oppositely under the $U(1)$ part of the gauge symmetry. This results in the vanishing of the one-loop beta function for the FI parameter. We suspect that the theory in fact admits an exactly marginal deformation on such loci. If this is the case, it would be nice to understand the corresponding exactly marginal deformations in other duality frames. The mid-point of the parameter space is a very special point as it is invariant under triality. The conformal manifold at this point could make an interesting study. We summarize the discussion of this subsection in figure \ref{fig:phasediagram}.

\section{Quivers}
\label{sec:general-quivers}\label{trialitywebs}

\begin{figure}[ht]
\centering
\includegraphics[scale=0.7]{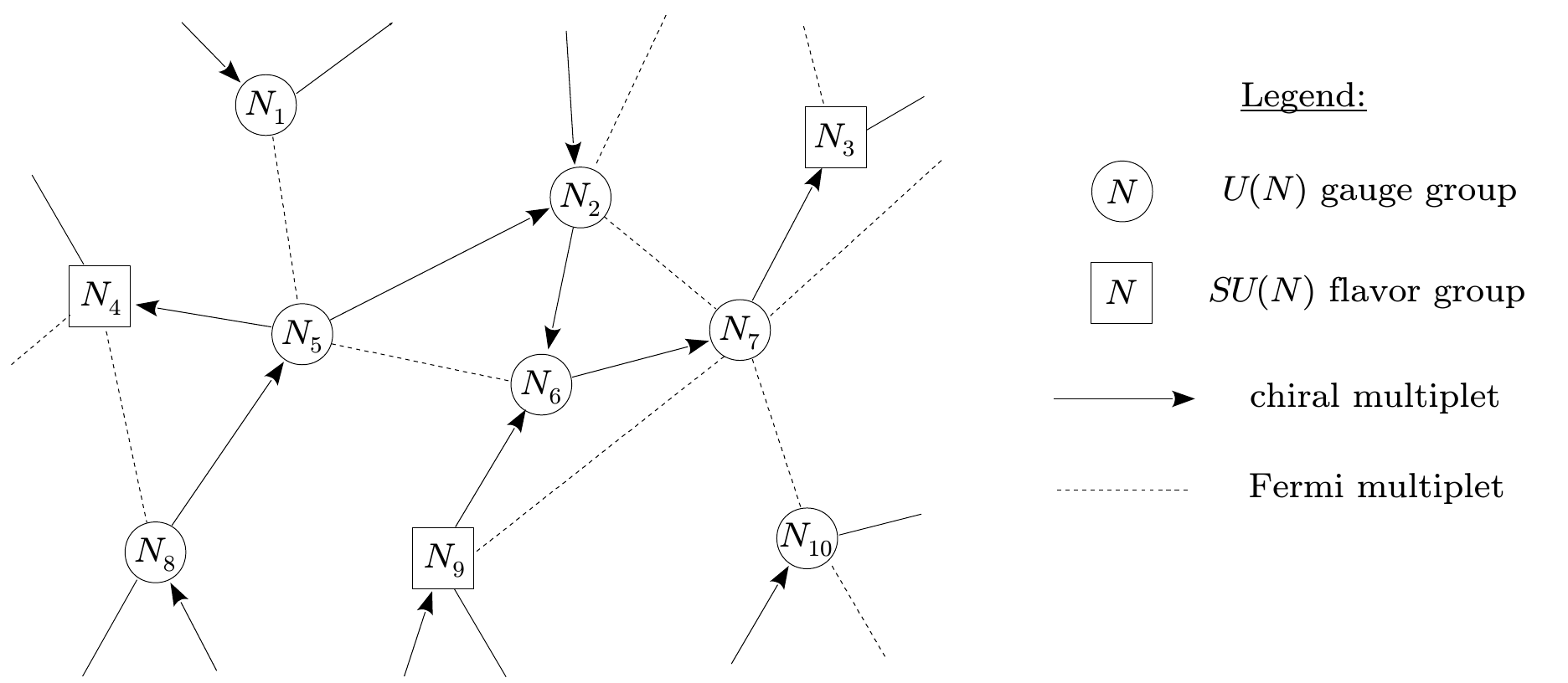}
\caption{An example of a general quiver.}
\label{fig:general-quiver}
\end{figure}

In this section we study triality actions on general 2d $\CN=(0,2)$ quiver gauge theories. An example of a general quiver is shown in figure \ref{fig:general-quiver}.
A cubic $J$-term superpotential is associated to all closed triangular loops in the quiver diagram. It is important that the representations of the chiral multiplets are compatible with such a superpotential. Moreover, we require every chiral multiplet to be part of a superpotential term. The orientation of the fermionic edge is automatically determined by the orientation of the bososnic edges.

For each gauge node
\footnote{To emphasize the rank of the gauge node $\circled i$, we sometimes use the notation  \includegraphics[scale=0.5]{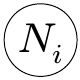}.}
 $\circled i$, let us define $\XX_i\equiv \{ \circled{j}: \circled j \rightarrow \circled i\}, \YY_i\equiv \{ \circled{j}: \circled j \leftarrow \circled i\}$ and $\ZZ_i\equiv \{ \circled{j}: \circled j$ - - $\circled i\}$. The cancellation of $SU(N_i)$ anomaly requires
\begin{equation}
 N_i \; = \; \Big(\sum_{\circled j \in \XX_i} N_j+ \sum_{\circled j \in \YY_i} N_j-\sum_{\circled j \in \ZZ_i} N_j\Big)/2 \,.
\label{local-anomaly-cancellation}
\end{equation}
This condition uniquely determines the ranks of gauge groups in terms of the ranks of flavor groups. In order to cancel the anomaly for the $U(1)_i$ part of the gauge node $\circled i$, we need to introduce Fermi multiplets $\Omega_\ell$ in representations $\det^{n_i^\ell}$ of $U(N_i)$. The $U(1)_i$ anomaly cancellation as well as the mixed anomaly cancellation between $U(1)_i$ and $U(1)_j$ require
\be
\sum_\ell n^\ell_in^\ell_j \; = \; 2\delta_{ij} - A_{ij} \,,
\label{abelian-anomaly-cancellation}
\ee
where $A_{ij}$ is the super-adjacency matrix of the quiver in which bosonic and fermionic edges contribute $+1$ and $-1$, respectively. It follows that if the gauge nodes form a tree, it should be of the ADE type because the vectors $\vec{n}_i$ define a root system. It is an interesting combinatorial exercise to classify all the graphs admitting solutions to  \eqref{local-anomaly-cancellation} and \eqref{abelian-anomaly-cancellation}. Note that, if we choose to gauge only the $SU(N)$ part of the gauge group then we do not need to worry about the condition \eqref{abelian-anomaly-cancellation}.

\subsection{The triality rules}
\begin{figure}[ht]
\centering
\includegraphics[scale=0.7]{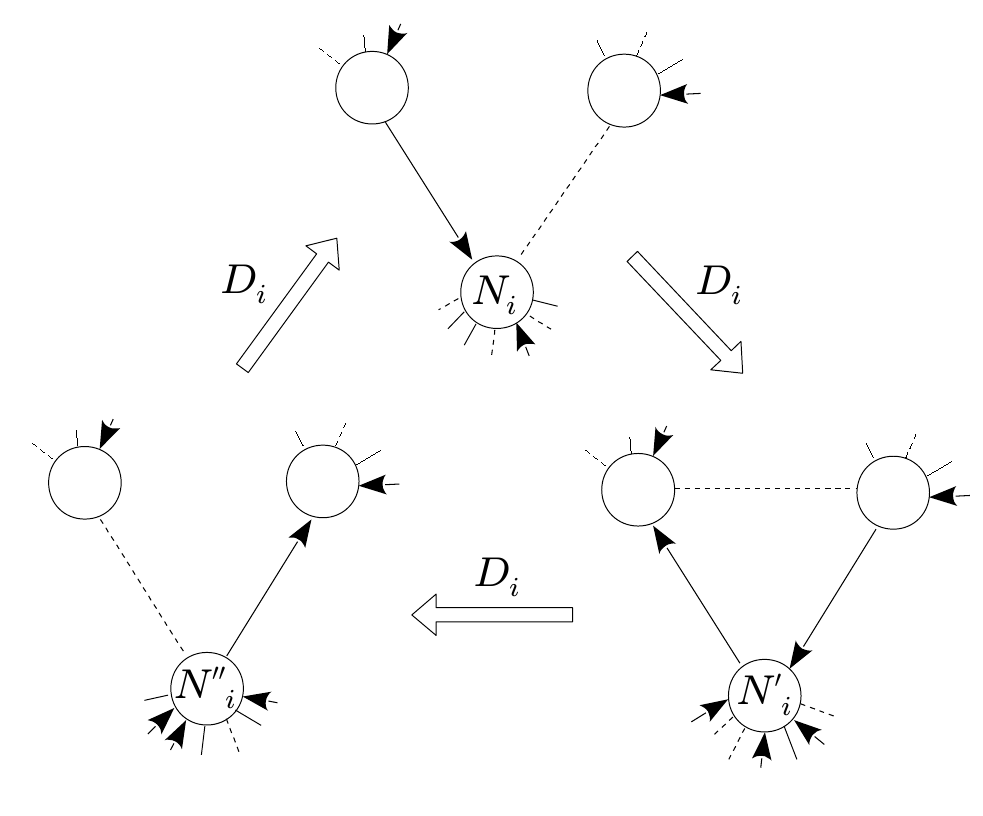}\qquad\qquad\includegraphics[scale=0.7]{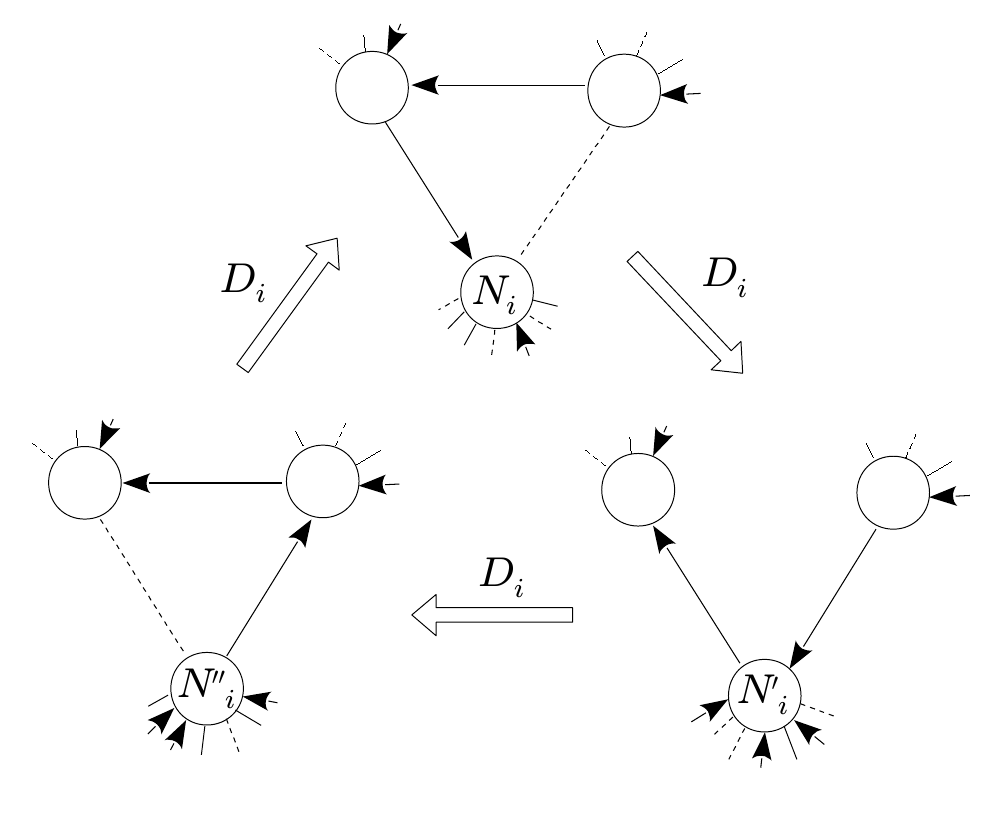}\\
\vspace{1cm}
\includegraphics[scale=0.7]{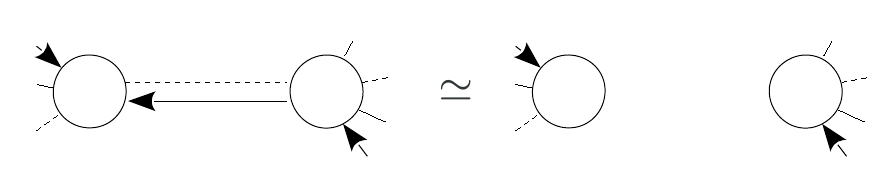}
\caption{Action of the triality $D_i$ associated to the gauge node $i$ and cancellation of fermi-bose pairs.}
\label{fig:local-triality}
\end{figure}

The triality of section \ref{triality} now acts on each individual node. The general transformation rules for a ``local'' triality at $\circled i$ are:
\begin{itemize}
\item Draw the same type of arrows from $\circled k \in \YY_i \cup \ZZ_i$ to all $\circled j \in \XX_i$ that connect $\circled k$ to the gauge node.
\item Change the connections to the gauge node s.t. all $\circled k \in \XX_i$ now belong to $\YY_i'$, all $\circled k \in \YY_i$ now belong to $\ZZ_i'$ and all $\circled k \in \ZZ_i$ now belong to $\XX_i'$.
\item The rank of new gauge group is $N_i'=\sum_{\circled j \in \XX_i} N_j- N_i$.
\item Cancel fermi-bose pairs.
\end{itemize}
These rules are illustrated in figure \ref{fig:local-triality}.
One can easily check that $N_i'$ automatically satisfies the new (primed) version of the condition \eqref{local-anomaly-cancellation}. The charges of $\Omega$ fermions transform as
\begin{equation}
\begin{array}{rcll}
\vec{n}'_i & = & \vec{n}_i \,, & \\
\vec{n}'_b & = & -\vec{n}_b-\vec{n}_i \,, & \quad {\circled{b}\rightarrow\circled{i}} \\
\vec{n}'_j & = & -\vec{n}_j \,, & \quad \text{\textit{all} other nodes}.
\end{array}
\label{n-vectors-transform}
\end{equation}
It is easy to check that the vectors $\vec{n}'_k$ satisfy the equations (\ref{abelian-anomaly-cancellation}) for the new quiver. In general,  performing the transformation \eqref{n-vectors-transform} thrice doesn't take us back to the original solution but rather produces a new solution to the condition \eqref{abelian-anomaly-cancellation}.

Now we show that local non-abelian anomalies are invariant under the local triality. Let the $SU(N)$ anomaly for $\circled j \in \XX_i$ be ${\cal A}_j$. After triality, the new edges from $\circled k \in \YY_i \cup \ZZ_i$ add  $\sum_{\circled k \in \YY} N_k -\sum_{\circled k \in \ZZ} N_k$. The contribution of the node $\circled i$ changes from $N_i$ to $N_i'$. All in all,
\be
{\cal A}_j' \; = \; {\cal A}_j+\sum_{\circled k \in \YY_i} N_k -\sum_{\circled k \in \ZZ_i} N_k +(N_i'-N_i) \; = \; {\cal A}_j \,.
\ee
The last equality follows from \eqref{local-anomaly-cancellation}. Similarly, one can verify the anomaly matching for $\circled j \in \YY_i$ and $\circled j \in \ZZ_i$. Matching of the equivariant index under the local triality is carried out in appendix \ref{appendix-theta}.

\subsection{Triality networks}

The computation of equivariant index demonstrates that the supersymmetry is dynamically broken if either $\sum_{\circled j \in \XX_i} N_j<N_i$ or  $\sum_{\circled j \in \YY_i} N_j<N_i$ for some $i$. This also means that, in such cases, the rank of the gauge group formally obtained by applying the triality rules is negative. As emphasized earlier, the condition \eqref{local-anomaly-cancellation} allows us to express the gauge group ranks uniquely in terms of the ranks of the flavor groups. The positivity conditions, in all duality frames, then carve out a polyhedron in the space of flavor group ranks.

\begin{figure}[ht]
\centering
\begin{tabular}{m{0.3\textwidth}m{0.15\textwidth}m{0.3\textwidth}}
\centering \includegraphics[scale=0.7]{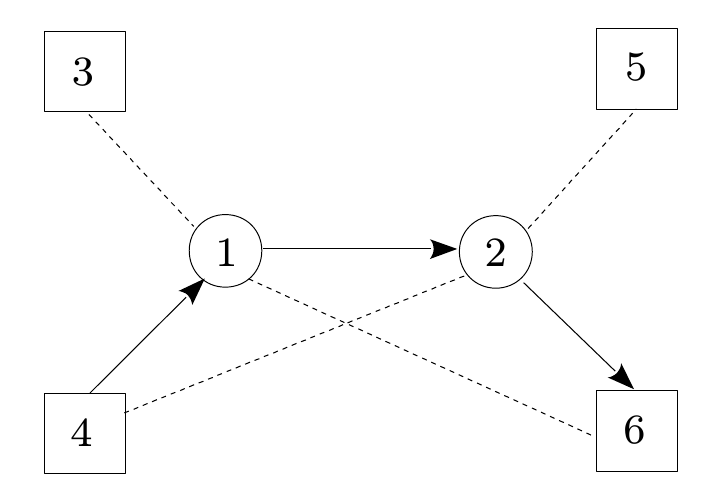} &
\centering $\stackrel{D_1}{\longrightarrow}$ &
\centering \includegraphics[scale=0.7]{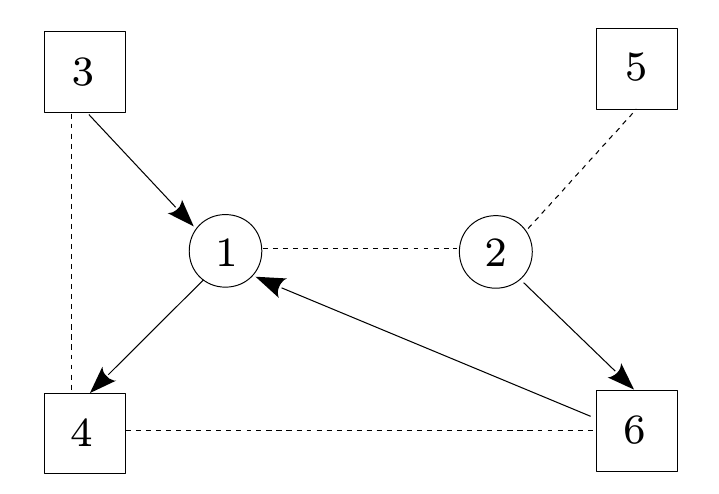}
\end{tabular}
\caption{Example of a quiver theory with dynamical supersymmetry breaking.}
\label{fig:bad-example}
\end{figure}

For some especially ``bad'' graphs the positivity conditions do not admit any solutions. In particular, a graph which has a dual with a gauge node $\circled i$ such that $\XX_i= \{\emptyset\}$ (or  $\YY_i= \{\emptyset\}$ or $\ZZ_i= \{\emptyset\}$) is bad. Consider the example in figure \ref{fig:bad-example}. Even though, the graph on the left appears to be innocent, its dual has a gauge node with no incoming arrows. In this description the quiver manifestly breaks supersymmetry for any values of the flavor group ranks. In fact, generic quiver graphs turn out to be bad in this sense. It will be interesting to come up with a combinatorial criterion for ``good'' graphs.

\begin{figure}[ht]
\centering
\includegraphics[scale=0.7]{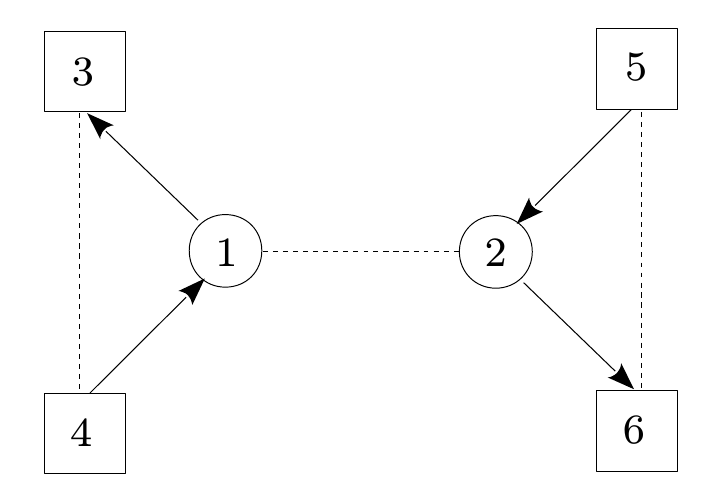}
\caption{An example of a theory with two gauge nodes that does not exhibit dynamical supersymmetry breaking.}
\label{fig:good-example}
\end{figure}

\begin{figure}[ht]
\centering
\includegraphics[scale=0.55]{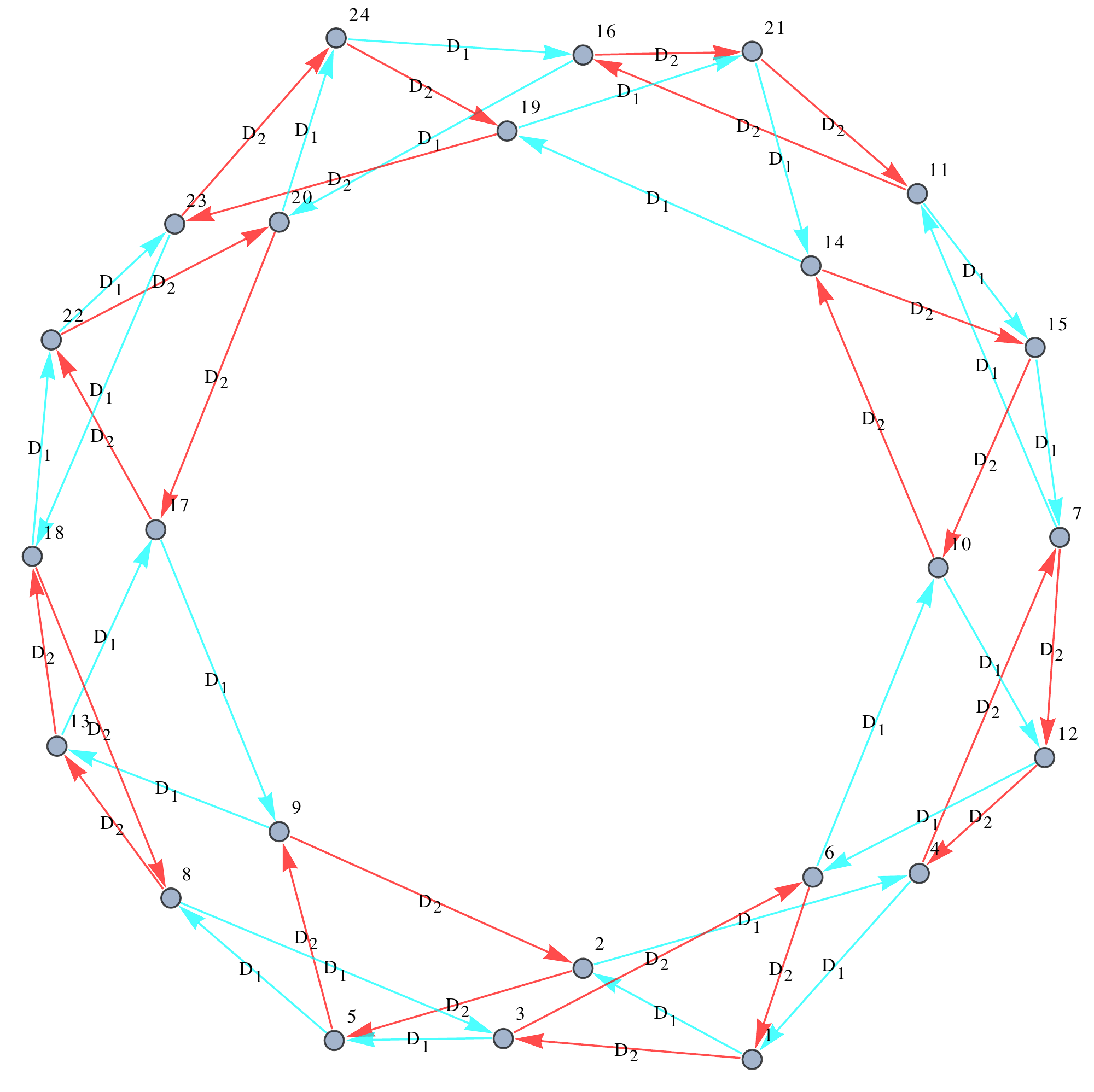}
\caption{The duality network of 24 dual theories generated by the actions of basic trialities $D_1$ and $D_2$ starting from the theory given by the quiver in figure \protect\ref{fig:good-example}.}
\label{fig:good-example-network}
\end{figure}

An example of a good graph with two nodes is shown in figure \ref{fig:good-example}. The conditions (\ref{abelian-anomaly-cancellation}) are met with $n_1=(1,1,0)$ and $n_2=(0,1,1)$. Applying trialities $D_1$ and $D_2$ we generate 24 quivers\footnote{This means counting quivers with marked gauge nodes. The network contains quivers which describe the same theory but differ by permutations of gauge labels.}. The triality network is displayed in figure \ref{fig:good-example-network}. Remarkably, other examples of two node quivers also have an isomorphic triality network. The positivity conditions amount to the bounds
\begin{equation}
 3N_i \; < \; \sum_{j=3}^6 N_j \,, \quad i=3,\ldots, 6 \,.
 \label{two-node-condition}
\end{equation}
They define the interior of an infinite cone over a tetrahedron in the 4-dimensional space of $(N_3,N_4,N_5,N_6)$.  On a face of the tetrahedron one of the gauge nodes has zero rank in a particular duality frame. Then, the theory effectively becomes identical to the theory with one gauge group, as in section \ref{triality}. And, each face of the tetrahedron plays the role of the triangular parameter space for the theory with one gauge node.

\begin{figure}[ht]
\centering
\includegraphics[scale=0.6]{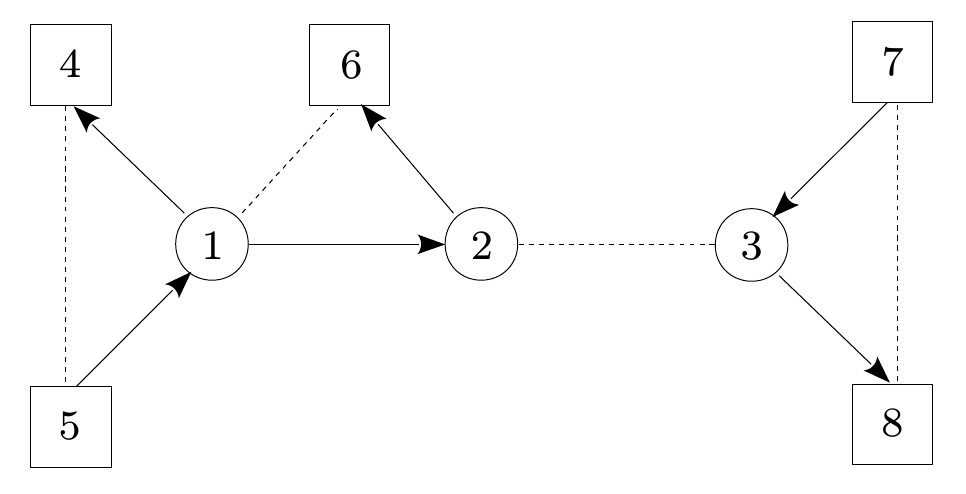}
\caption{An example of a theory with three gauge nodes without dynamical supersymmetry breaking.}
\label{fig:3node-example}
\end{figure}

Figure \ref{fig:3node-example} shows an example of a good graph with three gauge nodes. Its triality network consists of 330 quivers.


\section{Outlook}\label{outlook}

\begin{figure}[t]
\centering
\includegraphics[width=2.3in]{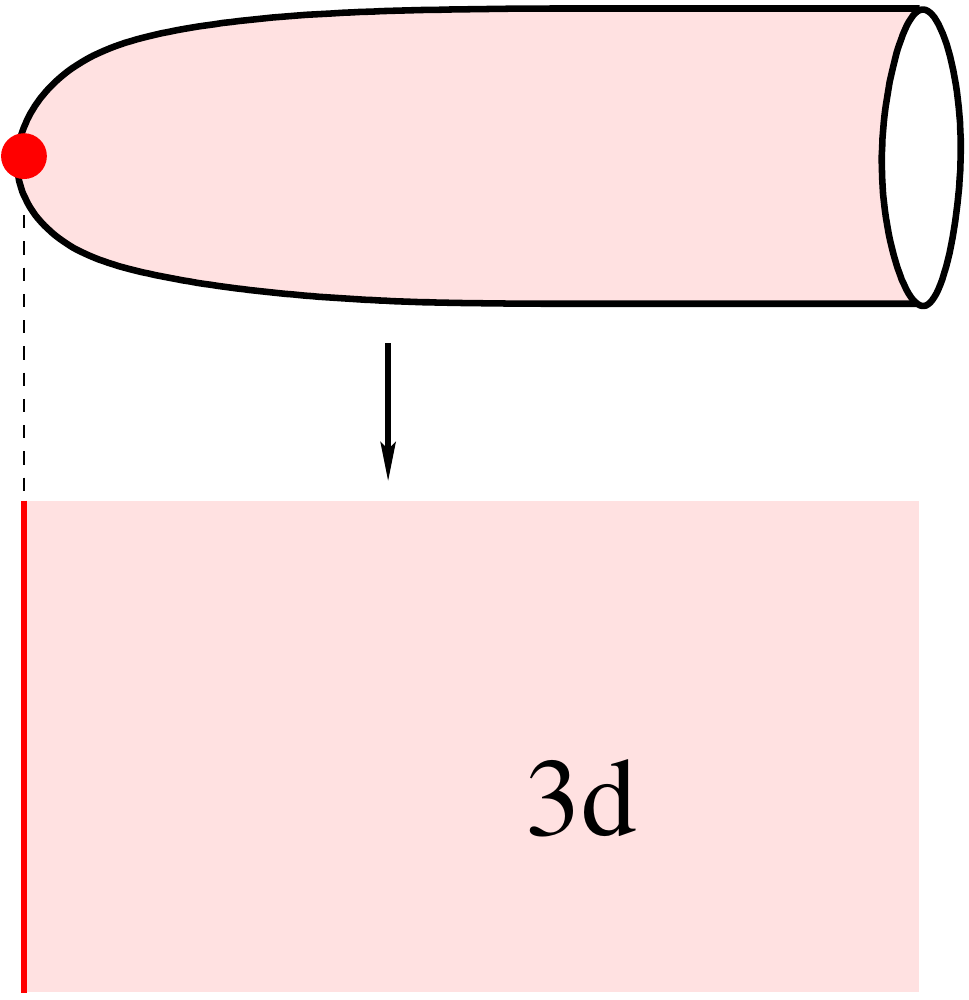}
\caption{A half-BPS surface operator in 4d $\CN=1$ gauge theory defines a half-BPS boundary condition in 3d $\CN=2$ theory.
Namely, consider a 4d $\CN=1$ gauge theory in space-time of the form $\R^2 \times D$
coupled to 2d $\CN=(0,2)$ theory (with symmetry group $G$) at the tip of the ``cigar'' $D \cong \R^2$. There are two ways to look at this system.
One, more obvious, is as a 4d-2d coupled system that describes a half-BPS surface operator in $\CN=1$ theory.
Another is based on a dimensional reduction on a circle (= the fiber of the ``cigar'' $D$).
This reduction gives a 3d $\CN=2$ gauge theory on a half-space, $\R_+ \times \R^2$, coupled to a 2d $\CN=(0,2)$ theory on the boundary.}
\label{fig:3dcigar}
\end{figure}

In this paper we have visited the uncharted landscape of 2d $\CN=(0,2)$ gauge theories.
The exploration motivates many questions. Below are some of the urgent ones.
\begin{itemize}
\item For certain special values of the ranks of the flavor symmetry groups, the $(0,2)$ theories could have exactly marginal deformations. It will be interesting to identify such points in the parameter space of quiver theories and study their conformal manifolds.
\item The type IIA brane construction of 2d $\CN=(0,2)$ gauge theories has been discussed in \cite{GarciaCompean:1998kh}. It is a natural question to understand the triality from the brane setup.
\item The answer to the previous question may provide the desired link between 4-manifolds and 2d $\CN=(0,2)$ theories $T[M_4]$. After all, the type IIA construction should be an $S^1$ compactification of the M5 brane setup. What does triality mean for 4-manifolds? We expect that it corresponds to the handle-slide moves.
\item The previous question, in particular the gluing of 4-manifolds, involves the study of half-BPS domain walls and boundary conditions in 3d $\CN=2$ theories that was recently initiated in \cite{Gadde:2013wq,Okazaki:2013kaa}. We expect the $(0,2)$ triality to play an important role in this study as well as in the study of surface operators in 4d $\CN=1$ gauge theories \cite{Gaiotto:2013sma} that also support $\CN=(0,2)$ supersymmetry and, via circle reduction, map to half-BPS boundary conditions in three dimensions, as illustrated in figure \ref{fig:3dcigar}.
\item The 4d Seiberg duality solves the Yang-Baxter equation, more accurately, the star-star equation \cite{Yamazaki:2013nra}. As a result, one associates a 1d quantum integrable system to every coupling independent observable of the gauge theory. We are tempted to speculate that the 2d $\CN=(0,2)$ triality may provide a solution to the ``tetrahedron equation'' which is associated to 2d quantum integrable systems \cite{Zamolodchikov:1981kf} (also see {\it e.g.} \cite{Bazhanov:2008rd} for recent work).
\item We have given examples of quivers that preserve supersymmetry as well as examples of quivers that break it dynamically. The former seem to be harder to construct. Therefore, it would be nice to come up with combinatorial criteria for the quivers with unbroken supersymmetry.
\end{itemize}

\acknowledgments{We would like to thank F.~Benini, N.~Bobev, N.~Seiberg, E.~Sharpe, M.~Shifman, A.~Vainshtein and E.~Witten for useful discussions.
The work of A.G. is supported in part by the John A. McCone fellowship and by DOE Grant DE-FG02-92-ER40701.
The work of S.G. is supported in part by DOE Grant DE-FG03-92-ER40701FG-02.
The work of P.P. is supported in part by the Sherman Fairchild scholarship and by NSF Grant PHY-1050729.
We would like to thank the Aspen Center for Physics and the 2013 Simons Workshop in Mathematics and Physics
for hospitality during various states of this work. The Aspen Center for Physics is supported in part
by the National Science Foundation under Grant No. PHYS-1066293.
Opinions and conclusions expressed here are those of the authors and do not necessarily reflect the views of funding agencies.}


\appendix

\section{Index for general quivers}
\label{appendix-theta}

\begin{figure}[ht]
\centering
\includegraphics[scale=0.7]{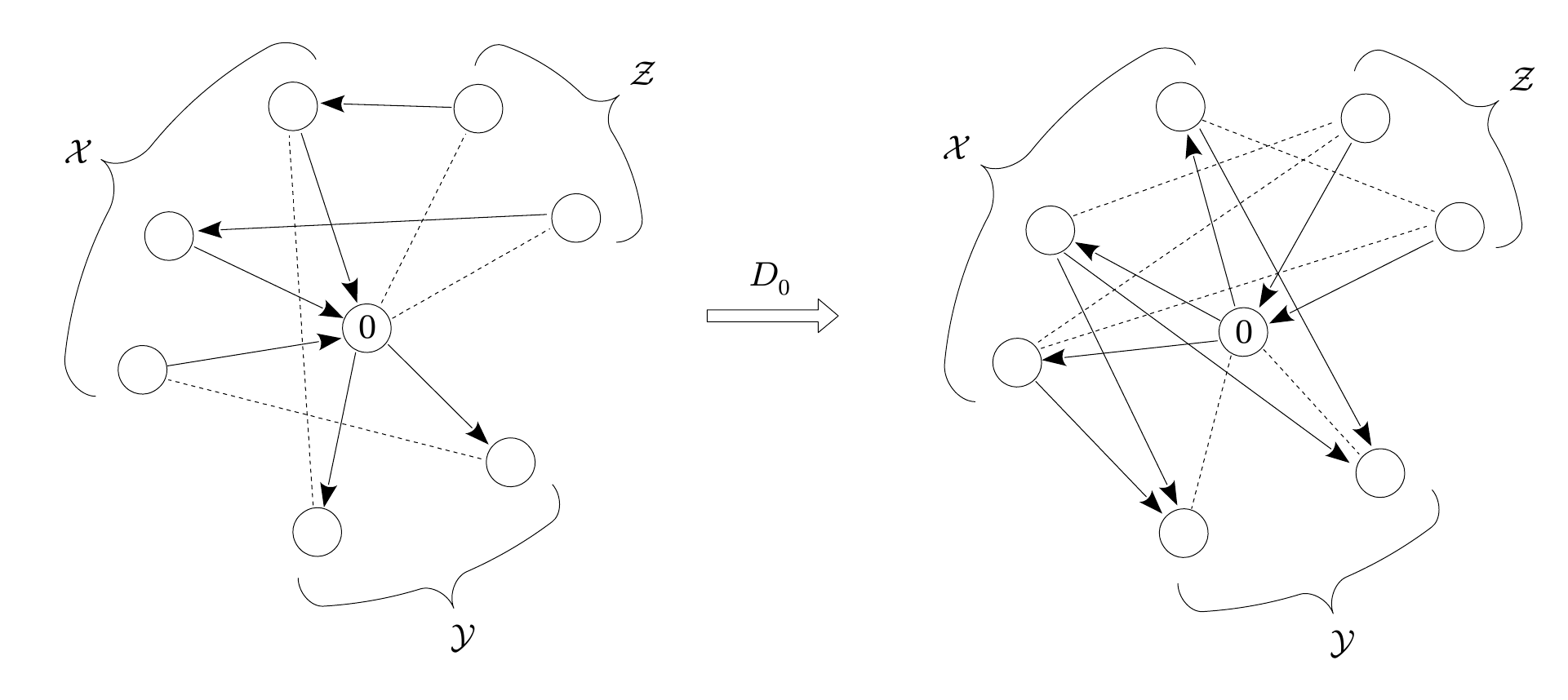}
\caption{Action of the triality $D_0$.}
\label{fig:index-identity}
\end{figure}

Here we show how the triality described in section \ref{sec:general-quivers} works at the level of index.
Let us pick a gauge node in a quiver and denote it by $0$. We separate nodes connected to the node $\circled{0}$ into 3 groups: nodes $\circled{a}\in \ZZ$ such that there is a Fermi multiplet $\circled{0}\text{- - -}\circled{a}$ in representation $\mathbf{N_a\otimes \bar N_0}$ and R-charge $R_a$, nodes $\circled{b}\in \XX$ such that there is a chiral multiplet $\circled{b}\rightarrow \circled{0}$ in representation $\mathbf{N_0\otimes \bar N_b}$ and R-charge $O_b$, and nodes $\circled{c}\in\YY$ such that there is a chiral multiplet $\circled{0}\rightarrow \circled{c}$ in representation $\mathbf{N_c\otimes \bar N_0}$ and R-charge $Q_c$ (See the left hand side of Fig. \ref{fig:index-identity} for an example). The anomaly cancellation requires that $2N_0=\sum_{\circled{b}\in\XX}N_b+\sum_{\circled{c}\in\YY}N_c-\sum_{\circled{a}\in\ZZ}N_a$. We will denote the fugacities for the corresponding gauge or flavor groups by $z_a^i$, $x_b^j$, $y_c^k$ and fugacities corresponding to the gauge node $\circled{0}$ by $\zeta^\alpha$. Then the part of the index for matter charged with respect to $U(N_0)$ and matter represented by lines between nodes $\circled{a},\circled{b},\circled{c}$ is given by
\begin{multline}
 \CI=\tilde{\CI}\int \prod_\alpha (q;q)_\infty^2\frac{d\zeta^\alpha}{\zeta^\alpha}\frac
 {
 \prod\limits_{\alpha\neq \beta}\theta(\zeta^\alpha/\zeta^\beta)
 \prod\limits_{\circled{a}\in\ZZ,\alpha,i}\theta(q^\frac{1+R_a}{2}z^i_a/\zeta^\alpha)
 }{
 \prod\limits_{\circled{b}\in\XX,\alpha,j}\theta(q^\frac{O_b}{2}\zeta^\alpha/x_b^j)
 \prod\limits_{\circled{c}\in\YY,\alpha,k}\theta(q^\frac{Q_c}{2}y^k_c/\zeta^\alpha)
 }
 \times
 \\
 \times
  \prod\limits_{\ell}
  \theta\left(q^\frac{1}{2}\left(\prod\limits_\alpha \zeta^\alpha\right)^{n_0^\ell} w_\ell \right)
\end{multline}
\begin{equation}
 \tilde{\CI}=\frac
 {
 \prod\limits_{\overset{\circled{c}\text{- - -} \circled{b}}{\circled{c}\in\YY,\circled{b}\in\XX}}\prod\limits_{j,k}\theta(q^\frac{2-O_b-Q_c}{2}x_b^j/y_c^k)
 }{
 \prod\limits_{\overset{\circled{c}\rightarrow \circled{a}}{\circled{c}\in\YY,\circled{a}\in\ZZ}}\prod\limits_{i,k}\theta(q^\frac{R_a-Q_c+1}{2}z_a^i/y_c^k)
 \prod\limits_{\overset{\circled{a}\rightarrow \circled{b}}{\circled{a}\in\ZZ,\circled{b}\in\XX}}\prod\limits_{i,k}\theta(q^\frac{1-R_a-O_b}{2}x_b^j/z_a^i)
 }
 \label{non-charged-matter}
\end{equation}
The integral can be performed using the residue theorem. The choice of poles can be specified by the injective map $\{z^\alpha\}\stackrel{\sigma}{\rightarrow} \{q^{-\frac{O_b}{2}}x^j_b\}$, $\text{Im}\,\sigma=\{q^{-\frac{O_b}{2}}\tilde{x}_b^r\}$.
\begin{multline}
 \CI=\tilde{\CI}
 \sum_{\{\tilde{x}_b^r\}\subset \{{x_b^j}\}}
 \frac{
 \prod\limits_{(r,b)\neq (r',b') }\theta(q^\frac{O_{b}-O_{b'}}{2}\tilde{x}^{r'}_{b'}/\tilde{x}^{r}_{b})
 \prod\limits_{r,i,a,b}\theta(q^\frac{1+R_a+O_b}{2}z^i_a/\tilde{x}^r_b)
 }{
 \prod\limits_{\tilde{x}^r_{b'}\neq x_b^j}\theta(q^\frac{O_b-O_{b'}}{2}\tilde{x}^r_{b'}/x_b^j)
 \prod\limits_{r,k,b,c}\theta(q^\frac{Q_c+O_b}{2}y^k_c/\tilde{x}^r_b)
 }
 \times
 \\
 \times
  \prod\limits_{\ell}\theta\left(q^\frac{1}{2}\left(\prod\limits_{b,r}q^{-\frac{O_b}{2}} \tilde{x}_b^r\right)^{n_0^\ell}w_\ell\right)
\end{multline}
After introducing the dual variables $\{\bar{x}^s_b\}=\{x^j_b\}\setminus \{\tilde{x}^r_b\}$ the index reads
\begin{multline}
 \CI=\tilde{\CI}
 \sum_{\{\bar{x}_b^s\}\subset \{{x_b^j}\}}
 \frac{
 \prod\limits_{(s,b)\neq (s',b') }\theta(q^\frac{O_{b}-O_{b'}}{2}\bar{x}^{s'}_{b'}/\bar{x}^{s}_{b})
 \prod\limits_{j,i,a,b}\theta(q^\frac{1+R_a+O_b}{2}z^i_a/{x}^j_b)
 \prod\limits_{s,k,b,c}\theta(q^\frac{Q_c+O_b}{2}y^k_c/\bar{x}^s_b)
 }{
 \prod\limits_{{x}^j_{b'}\neq \bar{x}_b^s}\theta(q^\frac{O_b-O_{b'}}{2}{x}^j_{b'}/\bar{x}_b^s)
  \prod\limits_{s,i,a,b}\theta(q^\frac{1+R_a+O_b}{2}z^i_a/\bar{x}^s_b)
 \prod\limits_{j,k,b,c}\theta(q^\frac{Q_c+O_b}{2}y^k_c/{x}^j_b)
 }
 \times
 \\
 \times
  \prod\limits_{\ell}\theta\left(q^\frac{1}{2}\left[\prod\limits_{b,j}q^{-\frac{O_b}{2}} {x}_b^j
  /\prod\limits_{b,s}\left(q^{-\frac{O_b}{2}}\bar{x}^s_b\right)\right]^{n_0^\ell}w_\ell\right)
\end{multline}
This can be represented as an integral over $N_0'=\sum_{\circled{b}\in\XX} N_b-N_0$ variables $\xi_\alpha$ which localizes to the poles given by the injective map $\{\xi^\alpha\}\stackrel{\sigma'}{\rightarrow} \{q^{\frac{\Delta-O_b}{2}}x^j_b\}$, $\text{Im}\,\sigma'=\{q^{\frac{\Delta-O_b}{2}}\bar{x}_b^r\}$:
\begin{multline}
 \CI=
 \frac{
  \prod\limits_{j,i,a,b}\theta(q^\frac{1+R_a+O_b}{2}z^i_a/{x}^j_b)
 }{
  \prod\limits_{j,k,b,c}\theta(q^\frac{Q_c+O_b}{2}y^k_c/{x}^j_b)
 }
 \,\tilde{\CI}
 \int \prod_\alpha (q;q)_\infty^2\frac{d\xi^\alpha}{\xi^\alpha}
 \frac{
 \prod\limits_{\alpha\neq \beta }\theta(\xi_\alpha/\xi_\beta)
 \prod\limits_{\alpha,k,c}\theta(q^\frac{Q_c+\Delta}{2}y^k_c/\xi^\alpha)
 }{
 \prod\limits_{\alpha,j,b}\theta(q^\frac{\Delta-O_{b}}{2}{x}^j_{b}/\xi^\alpha)
  \prod\limits_{\alpha,i,a}\theta(q^\frac{1-R_a-\Delta}{2}\xi^\alpha/z^i_a)
 }
 \times
 \\
 \times
  \prod\limits_{\ell}\theta\left(q^\frac{1}{2}\left(\prod\limits_{\alpha}\xi^\alpha\right)^{n_0^\ell}
  w_\ell^{-1}\left[\prod\limits_{j,b}x_b^j\right]^{-n_0^\ell}
  \right)
  \label{local-index-result}
\end{multline}
where
\begin{equation}
 \Delta=\frac{2\sum_{\circled{b}\in\XX} N_bO_b}{\sum_{\circled{b}\in\XX} N_b+\sum_{\circled{a}\in\ZZ} N_a-\sum_{\circled{c}\in\YY} N_c}.
\end{equation}
The integrand contains contributions from the following matter: Fermi multiplets $\circled{0}\text{- - -} \circled{c}$, $\circled{c}\in\YY$ with R-charges $Q_c+\Delta-1$, chiral multiplets $\circled{0}\rightarrow \circled{b}$, $b\in\XX$ with R-charges $\Delta-O_b$ and chiral multiplets $\circled{a}\rightarrow \circled{0}$, $\circled{a}\in\ZZ$ with R-charges $1-R_a-\Delta$. The new factors in front of the integral represent new bifundamental matter between nodes $\circled{a}\in\ZZ$, $\circled{b}\in\XX$ and $\circled{c}\in\YY$: $\circled{b}\text{- - -} \circled{a}$ and $\circled{b}\rightarrow \circled{c}$ for all pairs $(\circled{a},\circled{b})$ and $(\circled{a},\circled{c})$. The R-charges of these fields are consistent with superpotential given by the triangles where $\circled{0}$ is the third vertex. These contributions cancel with contributions to $\tilde{\cal I}$ from the original matter between pairs of nodes $(\circled{a},\circled{b})\in\ZZ\times\XX$ and $(\circled{a},\circled{c})\in\ZZ\times\YY$ given by (\ref{non-charged-matter}) (by using the identity $\theta(x)=\theta(q/x)$). Thus we verify that the theories related by the triality described in the section \ref{sec:general-quivers} have equal indices. The result (\ref{local-index-result}) is also consistent with the transformation rules (\ref{n-vectors-transform}).

In the rest of this section we will show that the identity between indices actually implies identity between central charges of the theories and their flavor anomalies. The gauge theories considered here have the property that the sum of all abelian gauge charges is even. This condition is related to the condition for the existence of spin structure i.e. $c_1(E)-c_1(TX)=0\;(\text{mod}\; 2)$  for a $(0,2)$ non-linear sigma model defined for a holomorphic bundle $E$ over $X$ (see e.g. \cite{KawaiMohri}). This implies the existence of a non-anomalous $\mathbb{Z}_2$ symmetry $B$. The index considered here has been tacitly twisted w.r.t. $B$. 

Using, 
\begin{equation}
 \theta(xq^\frac{1+R}{2})\overset{\hbar\rightarrow 0}{\sim} \exp\left\{
 -\frac{1}{2\hbar}\left[(\log x+\pi i)^2+\frac{\pi^2}{3}\right]
 +\frac{\hbar}{24}\left[-3R^2+1\right]
 -\frac{1}{2}R(\log x+\pi i)
 \right\}
 \label{theta-asympt}
\end{equation}
one can show that the index has the following asymptotics when $\hbar\rightarrow 0$:
\begin{equation}
 \CI\overset{\hbar\rightarrow 0}{\sim} \exp\left\{
 \frac{1}{2\hbar}\left[A+\frac{\pi^2}{3}(c_R-c_L)\right]
 +\frac{\hbar}{24}c_L
 \right\}
 \label{index-asympt}
\end{equation}
where
\begin{equation}
 c_L=\sum_{\text{Fermi mult. }\Psi}(1-3R_\Psi^2)+\sum_{\text{chiral mult. }\Phi}(3(R_\Phi-1)^2-1)
-2 \sum_{U(N)\text{ vector mult.} }N^2.
\end{equation}
is the left-moving central charge of the theory, $c_R-c_L=\Tr\gamma_3$ is the gravitational anomaly and $A$ is the anomaly polynomial. Namely,
\begin{equation}
A=\sum_{a,b}\left(\Tr\gamma_3 J_aJ_b\right)\log u_a\log u_b.
\end{equation}
where $u_a$ is the fugacity associated to the symmetry $J_a$ (symmetry $B$ has fugacity $-1$). Let us note that to obtain (\ref{index-asympt}) from (\ref{theta-asympt}) we used the fact that the mixed anomaly of R-symmetry with any other symmetry vanishes.

Therefore if two theories have equal indices they automatically have equal central chrages central charges and anomaly polynomials. Moreover, if the R-charges $R_a,O_b,Q_c$ of the original theory extremize the central charge, the R-charges of the dual theory extremize it too since they are related to the original R-charges through a linear transform.


\newpage

\bibliographystyle{JHEP_TD}
\bibliography{classH}

\end{document}